\documentclass[twocolumn, notitlepage,superscriptaddress]{revtex4-2}
\usepackage{blindtext}
\usepackage{graphicx}
\usepackage{dcolumn}
\usepackage{bm}
\usepackage{amssymb}
\usepackage{appendix}
\usepackage{url}
\usepackage{color}
\usepackage{xcolor}
\usepackage[T1]{fontenc}
\usepackage{mathrsfs,amsfonts,dsfont}
\usepackage{nccmath}
\usepackage{amstext}
\usepackage{mathtools}
\usepackage{enumitem}
\graphicspath{{graphics/}}
\usepackage{rotating} 
\usepackage[english]{babel}  
 
\usepackage{wrapfig}
\usepackage{lipsum} 

\usepackage[normalem]{ulem} 

\usepackage{braket}
\usepackage{bbm} 
\usepackage[colorlinks=true,citecolor=blue,linkcolor=blue]
{hyperref}

\usepackage{amsthm}
\newtheorem{theorem}{Theorem}
\newtheorem{lemma}{Lemma}

\begin{document}

\title{Multiplexed multipartite quantum repeater rates in the stationary regime}

\author{Julia A. Kunzelmann}
\thanks{These two authors contributed equally}

\affiliation{Institute for Theoretical Physics III, Heinrich Heine University Düsseldorf, D-40225 Düsseldorf, Germany}

\author{Anton Trushechkin}

\thanks{These two authors contributed equally}

\affiliation{Institute for Theoretical Physics III, Heinrich Heine University Düsseldorf, D-40225 Düsseldorf, Germany}

\affiliation{Steklov Mathematical Institute of Russian Academy of Sciences, Gubkina Str. 8, 119991 Moscow, Russia}

\author{Nikolai Wyderka}

\affiliation{Institute for Theoretical Physics III, Heinrich Heine University Düsseldorf, D-40225 Düsseldorf, Germany}

\author{Hermann Kampermann}

\affiliation{Institute for Theoretical Physics III, Heinrich Heine University Düsseldorf, D-40225 Düsseldorf, Germany}

\author{Dagmar Bruß}

\affiliation{Institute for Theoretical Physics III, Heinrich Heine University Düsseldorf, D-40225 Düsseldorf, Germany}

\begin{abstract}
    Multipartite quantum repeaters play an important role in quantum communication networks enabling the transmission of quantum information over larger distances. 
    To increase the rates for multipartite entanglement distribution, multiplexing of quantum memories is included. 
    Understanding the limitations of achievable rates in the stationary regime for different network sizes is a fundamental step to comprehend scalability of quantum networks. 
    This work investigates the behavior of the multipartite quantum repeater rate (i.e., the number of GHZ states generated per round and per memory) in the stationary regime in multipartite star graphs with a single central multipartite quantum repeater including multiplexing using Markov chains. 
    We derive a closed-form expression for the stationary rate depending on the network size.
    We support our results with numerical simulations. 
    Further, we show that the rate saturates for large number of memories. On an abstract level, the mathematical description is equivalent to quantum repeater chains between two parties. Therefore, our results also apply to those setups. 
\end{abstract}
\maketitle

\section{Introduction}

Quantum communication over large distances in quantum networks relies on the availability of entangled quantum pairs between end nodes. 
By increasing the distance between two nodes, the success probability for distributing entangled quantum states decreases~\cite{Repeater}. 
To overcome this problem, quantum repeaters have been suggested~\cite{Repeater1, QuantumRepeater2, Hartmann}. 
By performing Bell state measurements in a central quantum repeater, entanglement between two distant parties can be generated. 
Analogously, in the multipartite setup, a GHZ measurement can be performed to entangle all $n$ parties that are connected with the multipartite quantum repeater~\cite{Avis_2023}.  
To further increase the rate of entanglement distribution in quantum networks, multiplexing can be introduced to the quantum repeater~\cite{Multi_Original}, where each party has $m$ parallel quantum channels with the repeater. 
Instead of sending one qubit to the repeater, $m$ qubits can be sent in parallel. 
Consequently, up to $m$ entangling operations can be performed in parallel at the central station, such that the entanglement distribution rate is increased. 
This has been analyzed theoretically for the bipartite~\cite{Multiplexing} as well as the multipartite setup~\cite{kunzelmann_paper}. Since multiplexing is related to an increased cost of resources, understanding the scaling of such multiplexed multipartite quantum repeaters with respect to the number of parties and memories per party is of great interest. 
\\
\par
A repeater chain of $n$ segments with $n-1$ repeater stations between two parties was considered in Ref.~\cite{Shchukin2019}. By using Markov chains, the authors calculated the average waiting times and, based on this, the transmission rates. 
In Ref.~\cite{Khatri}, also the average waiting time was analyzed for evaluating entanglement distribution in quantum networks. 
The entanglement generation rate was analyzed in Ref.~\cite{Vinay}. 
In Ref.~\cite{Kamin}, the secret key rate for repeater chains with more segments was computed. 
The authors additionally considered more realistic setups, including memory cutoffs and repeater chains running in parallel.
Bipartite quantum repeaters with multiplexing were studied in Ref.~\cite{Collins, Jiang, Ravazi}.

In this paper, we investigate multipartite quantum repeaters that connect more than two parties based on the setup shown in Fig.~\ref{fig:Setup}. 
Here, one multipartite quantum repeater is placed between $n$ parties, that are connected to a central station. 
Each party has $m$ memories available at the multipartite quantum repeater. 
These memories allow the parties to send $m$ qubits in parallel. 
The purpose of the multipartite quantum repeater is to generate Greenberger-Horne-Zeilinger (GHZ) multipartite entangled states. 
As a figure of merit, we consider the average number of GHZ states generated per round and per memory, called ``multipartite repeater rate''. In the following, we simply denote it as repeater rate.

In contrast to previous works, we are not interested in the average waiting time for generating the desired state, starting from the initial state of empty memories, but in the repeater rate of the stationary (i.e., long-term) regime. The difference with the scenario of Ref.~\cite{Kamin} is that, after a successful GHZ measurement, we do not empty all multiplexed memories, but keep them for future rounds. In other words, the next round does not start again from the initial state, but starts from a state where some memories can be already filled. Then we obtain a random process (a Markov chain) and are interested in its stationary regime, namely in stationary repeater rate.

The problem of finding the stationary state of a Markov chain is reduced to the solution of a system of linear equations. It can be solved numerically, but it is preferable to have an analytic solution for understanding the influences of various parameters on the performance of such systems. Unfortunately, this system is intractable analytically, even for a moderate number of parties and memories. 
So, suitable approximations are required, which we use to derive explicit formulas that give a very good approximation for the repeater rate for an arbitrary number of parties and memories.

It is worth noting that, on the considered idealized level of description, our scenario is mathematically equivalent to a transmission line between two participants, Alice and Bob, with $n-1$ intermediate quantum repeaters that split the transmission line into $n$ segments and generate Bell pairs on each segment. Thus, though we will focus on the case of a multipartite quantum repeater between $n$ participants, our results can be applied also to a chain of $n-1$ quantum repeaters between two participants.
\\
\par
Our paper is structured as follows. In Sec.~\ref{sec:Setup}, we introduce the $n$-partite quantum repeater with multiplexing and its description via Markov chains. In Sec.~\ref{sec:nomulti}, we analyze the setup without multiplexing and show the equivalence to the setup of a bipartite repeater chain.  
We move on to the generalized multipartite quantum repeater including multiplexing in Sec.~\ref{sec:Generalization}. 
We give bounds on the repeater rate in the multipartite repeater setup, including multiplexing. 
We further discuss the effect of memories on the repeater rate and the dependency on the number of parties as well as the number of memories in larger networks. In Sec.~\ref{sec:summary}, we present our conclusion and outlook.

\section{The setup} 
\label{sec:Setup}

\subsection{Physical description}

\begin{figure}
    \centering
    \includegraphics[scale=0.2]{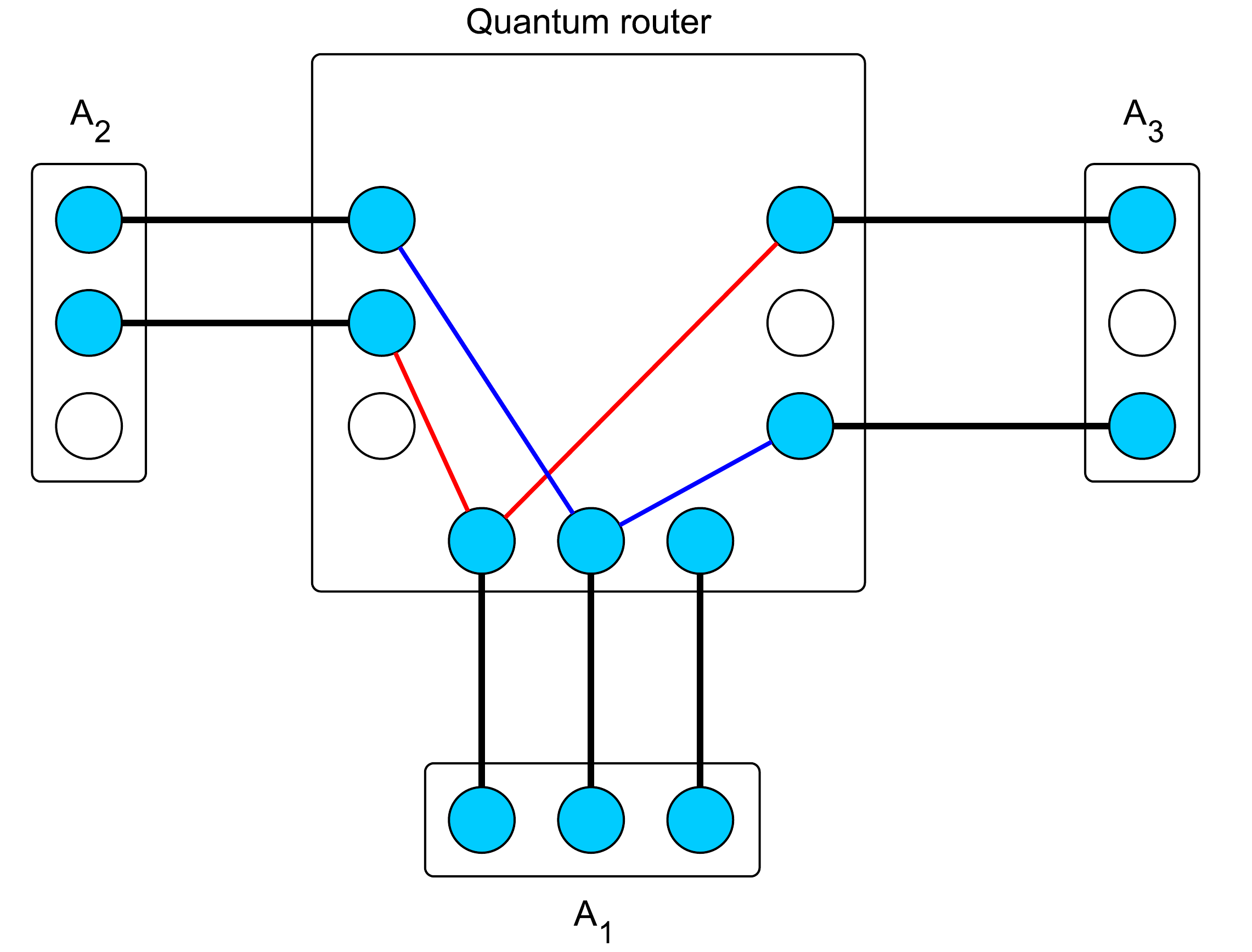}
    \caption{Setup of a tripartite multipartite quantum repeater with $n=3$ parties and $m=3$ memories per party adapted from~\cite{kunzelmann_paper}. The blue and white vertices (balls) correspond to empty and filled quantum memory cells, respectively. The edges between the parties and the multipartite quantum repeater (depicted in black) correspond to established Bell pairs. Edges inside the multipartite quantum repeater (depicted in red and blue) correspond to possible combinations allowing two GHZ measurements.}
    \label{fig:Setup}
\end{figure}

We consider a star network with one central multipartite quantum repeater and $n$ parties around it~\cite{kunzelmann_paper} (see Fig.~\ref{fig:Setup}). 
Each party has $m\geq1$ sources producing Bell pairs. 
We will refer to the case $m>1$ as multiplexing. 
Additionally, each party has $m$ quantum memories in the multipartite quantum repeater to store arriving qubits~\cite{Multi_Original}. For simplicity, we assume that the memories are perfect and have infinite storage time.
Each source sends one qubit of the Bell pair via a quantum channel to the multipartite quantum repeater per unit of time (round).
The qubit is stored and heralded if it arrives successfully (with probability $p$). If each party has at least $l$ filled memories, then the parties can perform $l$ GHZ measurements. Fig.~\ref{fig:Setup} shows a configuration allowing two GHZ measurements. 

We consider the case of no restriction on the coupling of memory cells inside the multipartite repeater, i.e., if each party has at least one filled memory, a GHZ measurement can be performed independently of the positions of the filled memories in the memory stacks of each party. The case of restrictions was analyzed in Ref.~\cite{kunzelmann_paper}.

Recall that the $n$-partite GHZ measurement is the measurement corresponding to the orthonormal basis 
\begin{equation}
    \{X_1^{b_1}\ldots X_{n-1}^{b_{n-1}}Z_n^{b_n}
    \ket{\rm GHZ}\},
\end{equation}
$b_1,\ldots,b_n\in\{0,1\}$,
where
\begin{equation}
    \ket{\rm GHZ}=\frac1{\sqrt2}(
    \ket{0}^{\otimes n}
    +\ket{1}^{\otimes n})
\end{equation}
and $X_i$ and $Z_i$ are the Pauli operators acting on the $i$th qubit. Each state from this basis (i.e., a postmeasurement state) is local-unitary equivalent to the GHZ state, and the local unitaries are determined by the announced measurement outcome. We assume that the GHZ measurements are perfect.

The GHZ states are then used in an application, e.g., in a conference key agreement protocol~\cite{CKA}. Thus, 
memories that were included in a measurement are emptied again, see Fig.~\ref{Fig:transition}. All other memories remain untouched for the next round.
In the following, qubits from a Bell pair are only sent to memories that are not filled from a previous round.
\par 
The goal of the protocol is to perform as many GHZ measurements $l$ as possible per round and consequently maximize the rate in the stationary regime. More precisely, the number of GHZ measurements is a random variable $L$ (all random variables will be denoted as capital Latin letters, and their possible values will be denoted as the corresponding small letters) because the storage process is probabilistic. We want to maximize its average value $\langle L\rangle$ in the stationary regime. 

\subsection{States of the multipartite quantum repeater and transitions}
\label{sec:trans}

Let us describe the set of configurations of the memory cells in the multipartite quantum repeater. The multipartite repeater has $m$ qubit memory cells for each party, i.e., $nm$ memory cells in total. Each party also has $m$ memory cells, which are in one-to-one correspondence with the corresponding multipartite repeater's memory cells, see Fig.~\ref{fig:Setup}. Thus, each party's memory cell is assigned to its own multipartite repeater memory cell and one can speak about pairs of memory cells. Since the memories are filled by pairs, we will use the notions of pairs of memory cells or single memory cells (from the party or repeater side) interchangeably.

Thus, the configuration of memories in the whole system is uniquely defined by the memory configuration in the multipartite repeater. The configuration of the pairs of memories for the $i$th party can be defined as a binary vector $\vec{a}_i = \left( a_{i1}, a_{i2}, \hdots , a_{im} \right)$ with length $m$. Here, $a_{ij}=0$ and $a_{ij}=1$ correspond to an empty and filled $i$th pair of memories, respectively. 
The total memory configuration is thus given by $\mathbf{a} = ( \Vec{a}_1, \Vec{a}_{2}, \hdots, \Vec{a}_{n})\in\{0,1\}^{nm}$. 

Since the positions of the filled memories are unimportant for the possibility of the GHZ measurement, we are actually interested in the number of filled memories $|\vec a_i|$ (here, $|\cdot|$ denotes the Hamming weight of a binary vector, i.e., the number of ones) for each party rather than the vectors $\vec a_i$ themselves. This can be used in simulations to reduce the configuration space, and we will use this more economic description in Sec.~\ref{sec:Generalization}.  However, in this section, it will be conceptually more convenient to use the vectors $\vec a_i$.

\begin{figure}
    \centering
    \includegraphics[scale=0.255]{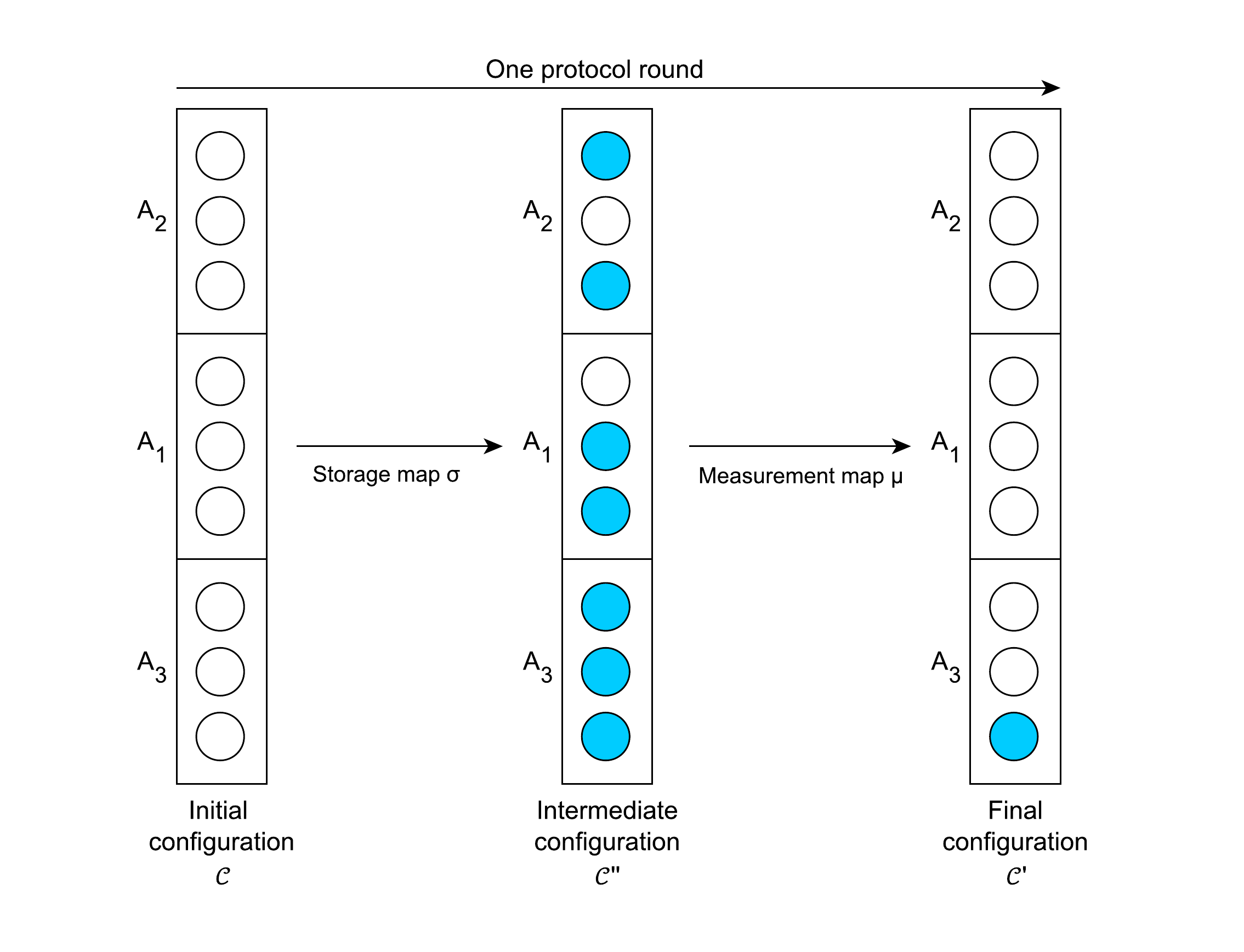}
    \caption{Schematic representation of a protocol round analogous to~\cite{kunzelmann_paper, Multiplexing}. Shown are the memory configurations for a multipartite quantum repeater with $n=3$ parties, each with $m=3$ memories. The transition of the memory configurations consists of the two processes of sending/storing (map $\sigma$) and measuring (map $\mu$).}
    \label{Fig:transition}
\end{figure}

Each round of the multipartite quantum repeater includes two steps, depicted in Fig.~\ref{Fig:transition}. 
The first step is the storage step. A Bell state source is assigned to each of the $nm$ pairs of memories between the parties and the multipartite repeater. During the storage step, for each empty pair of memories, the corresponding Bell state source tries to establish a Bell pair and store it in the pair of memories. The success probability of this single event is $p$. Possible reasons for the failure of Bell-state generation are discussed, e.g.,  in Ref.~\cite{Bernardes2011}. Note that the success of a Bell state generation is considered independently for each link.

The second step is called ``measurement'' which corresponds to emptying some memories due to the GHZ measurement. Namely, if each party has at least $l$  filled memories and one party has exactly $l$ filled memories, then $l$ GHZ measurements are performed and $l$ memories are emptied for each party. The precise positions of the emptied memories are unimportant. For definiteness, let us assume that each party tries to empty memories with higher indices first.

To answer the question of how many GHZ measurements can be performed per round in the stationary regime, we will employ the theory of Markov chains. 

\subsection{Description via Markov chains} 

Let $\pi_{\mathbf a}$ denote the probability for the system to be in the  configuration $\mathbf a\in\{0,1\}^{nm}$ in some moment of time. The probability distribution $\{\pi_{\mathbf a}\}_{\mathbf a\in\{0,1\}^{nm}}$ can be represented as a vector $\pi$ of length $nm$. 
Then the two steps described in the previous subsection correspond to $2^{nm}\otimes 2^{nm}$ transition matrices $\sigma$ (for the storage step) and $\mu$ (for the measurement step) between the configurations. 
The (stochastic) transition matrix $T$ of one ``working cycle'', or round, is then given by
\begin{align}
    T= \mu \sigma .  \label{eq:T}
\end{align}
A matrix element $T_{\mathbf{a}'\mathbf{a}}$ gives the probability of changing the multipartite quantum repeater's state from any configuration $\mathbf{a}$ to any configuration $\mathbf{a}'$ (i.e., the column contains the actual configuration and the row the future configuration).

As an example, consider the case of $n= 2$ and $m=1$, i.e., two participants, each with a single memory. As soon as both memories are filled with a quantum bit, both memories are emptied by performing a Bell state measurement. 
In this case, $\vec a_{1}$ and $\vec a_2$, are simply single bits and we can write $\mathbf a=(a_1,a_2)$.

\begin{figure}
    \centering
    \includegraphics[scale=0.2]{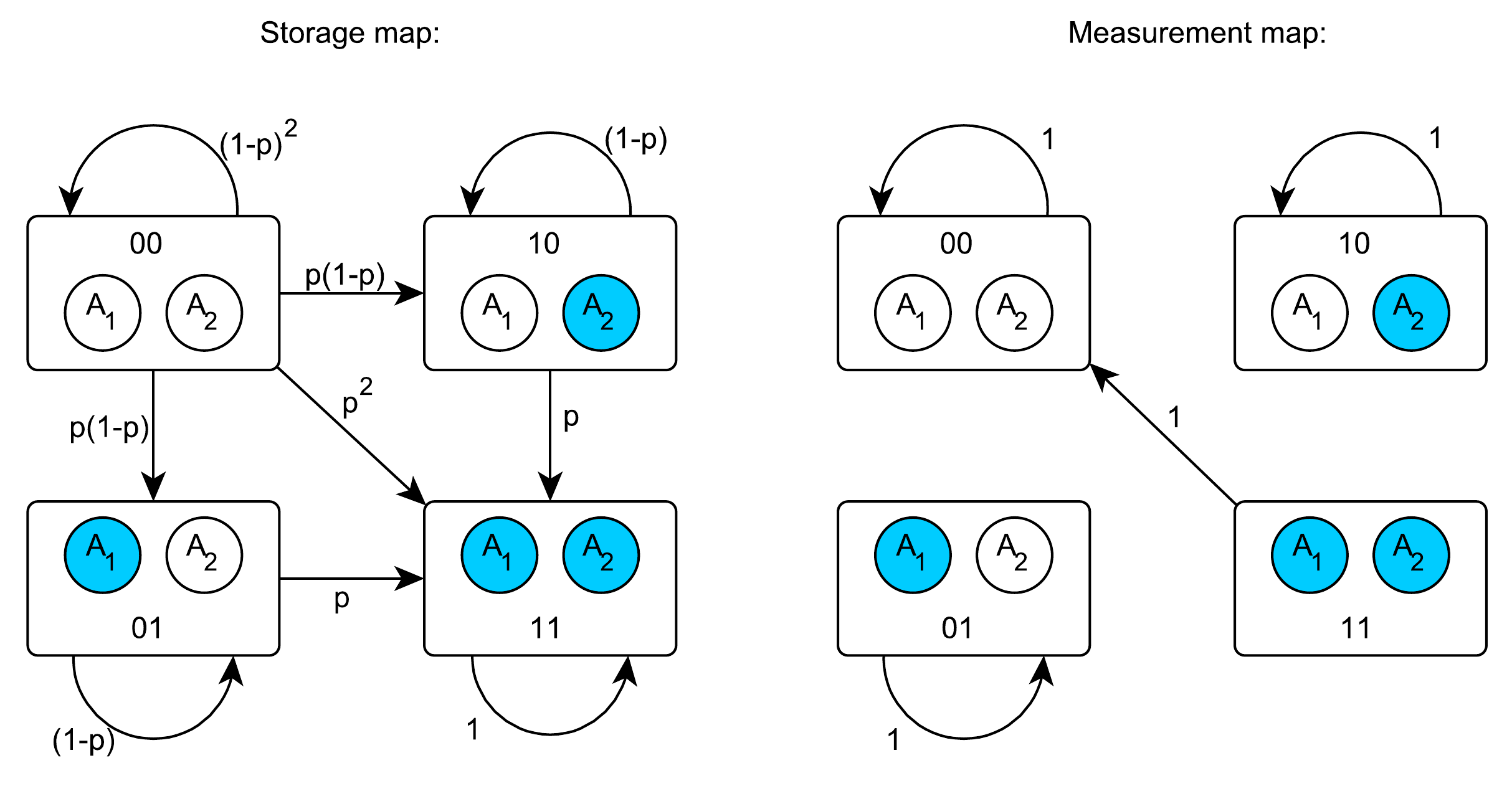}
    \caption{Graph representation of the storage map $\sigma$ (left) and the measurement map $\mu$ (right) of a bipartite quantum repeater with one memory per party. The vertices correspond to configurations, and the edges correspond to possible transitions between them. Along the edges, the transition probabilities depending on the success probability $p$ are given. The Bell measurement is considered to be perfect. }
    \label{fig:MCex}
\end{figure}

Such a Markov chain has four different configurations: $\mathbf{a}=(0,0)$ (both memories in the repeater are empty), $\mathbf{a}=(0,1)$ and $\mathbf{a}=(1,0)$ (one memory is filled and one is empty) and $\mathbf{a}=(1,1)$ (both memories are filled). The storage map $\sigma^{(1)}$ for one party has the form
,
\begin{align}
    \sigma^{(1)} = \begin{pmatrix}
    1-p & 0 \\
    p & 1
  \end{pmatrix}. 
\end{align}
Here, the first and second rows/columns correspond to $a_i=0$ (empty memory) and $a_i=1$ (filled memory), respectively. The first column indicates that an empty memory remains empty with probability $1-p$ and becomes filled with probability $p$. The second column means that the filled memory remains filled with probability 1 in this stage. 
The total storage map $\sigma$ is obtained via the tensor product:
\begin{align}
    \sigma = \sigma^{(1)} \otimes \sigma^{(1)} 
    = \begin{pmatrix}
        (1-p)^2 & 0 & 0 & 0 \\
        p(1-p) & 1-p & 0 & 0 \\
        p(1-p) & 0 & 1-p & 0 \\
        p^2 & p & p & 1
  \end{pmatrix},  
\end{align}
where the rows and columns (corresponding the bipartite configurations $\mathbf a$) are ordered in the usual way: $(0,0)$, $(0,1)$, $(1,0)$, $(1,1)$.
The measurement map $\mu$ is as follows:
\begin{align}
    \mu = \begin{pmatrix}
        1 & 0 & 0 & 1 \\
        0 & 1 & 0 & 0 \\
        0 & 0 & 1 & 0 \\
        0 & 0 & 0 & 0
    \end{pmatrix}.
\end{align}
That is, the memory configuration changes in this step only if both memories are filled and a Bell state measurement can be performed. 
A graph representation of both maps is shown in Fig.~\ref{fig:MCex}.
The transition matrix $T$ is then given by
\begin{equation}
    T = \mu \sigma   
    = \begin{pmatrix}
    (1-p)^2+p² & p & p & 1 \\
    p(1-p) & 1-p & 0 & 0 \\
    p(1-p) & 0 & 1-p & 0 \\
    0 & 0 & 0 & 0
  \end{pmatrix}.
\end{equation}
\par
Returning from our example, the stationary distribution $\pi^*$ of a Markov chain is a distribution $\pi$ (represented as a column vector) that does not change when the transition matrix $T$ is applied:
\begin{align}
\label{eq:Tpi}
    \pi^* = T  \pi^*.
\end{align}
The stationary distribution can also be calculated by taking the limit of $T^s$ with infinitely many rounds $s$ and applying it to an initial distribution $\pi^{(0)}$:
\begin{align}
\label{eq:stationary_distribution}
    \pi^* = \lim_{s \rightarrow \infty} T^s \pi^{(0)}.
\end{align}
Note that $\pi^*$ is the stationary probability distribution for memory configurations at the end of a round, i.e., after a measurement step. The number of GHZ state measurements in any round is specified by the configuration before this step, i.e., after the storage map. So, to get the average number of GHZ state measurements, we need to apply the storage map, i.e., we need to consider the vector $\sigma\pi^*$. 

The number of GHZ measurements in the stationary regime as a random variable is denoted by $L$. The average value of the number of the GHZ measurement $\langle L\rangle$ is completely determined by $\sigma\pi^*$: 
\begin{equation}
\label{eq:ell}
    \langle L\rangle
    =
    \sum_{\mathbf a\in\{0,1\}^{nm}}
    (\sigma\pi^*)_{\mathbf a}
    \min_{i=1,\ldots,n}
    |\vec a_i|,
\end{equation}
where $(\sigma\pi^*)_{\mathbf a}$ denotes the entry of the vector $\sigma\pi^*$ and, recall, $|\cdot|$ denotes the Hamming weight of the vector. The minimal Hamming weight over all parties is the number of the GHZ measurements $l$ for a given configuration. From that, the asymptotic repeater rate per memory can be calculated as
\begin{align}
\label{eq:Router_rate}
    R_\infty = \frac{\langle  L \rangle}{m} 
\end{align}
with $m$ being the number of memories per party.

The Markov chain that represents the transitions of a multipartite quantum repeater during one round is irreducible. 
A Markov chain is irreducible if 
any configuration can reach all other configurations (within a finite number of rounds). 
This is true in the setup of the multipartite quantum repeater, as any number of memories can be filled in every round, and the configuration with all memories filled can always be reached, allowing them to be emptied again. Therefore, it is always possible to reach any configuration $\mathbf{a}'$ from any configuration $\mathbf{a}$ within a finite number of rounds. 
The stationary distribution of an irreducible Markov chain is unique, i.e., it is independent of the initial distribution $\pi^{(0)}$. A natural choice for $\pi^{(0)}$ is the case of all memories being empty.

\section{multipartite quantum repeater without multiplexing} 
\label{sec:nomulti}

First, we consider a multipartite quantum repeater with a single memory per party, i.e., without multiplexing. 
In this case, $\mathbf a=(a_1,\ldots,a_n)$, where $a_i$ are bits corresponding to the states of the single memories of each party. A GHZ measurement is possible only in the configuration $\mathbf a=(1,\ldots,1)$, i.e., the memory for each party is filled. The average number of GHZ measurements per round, or, equivalently, for this case, the probability of performing a single GHZ measurement in the stationary regime, is given by
\begin{equation} 
    \label{eq:Prob_L1}
    \langle L_1 \rangle  = \Pr (L_1=1)   
   =(\sigma\pi^*)_{(1,\ldots,1)},
\end{equation}
where the subscript 1 in the notation $L_1$ denotes that we consider the single-memory (per party) case.  It turns out that
\begin{equation}
\label{eq:Result_no_multiplexing}
   \langle L_1 \rangle  
   = \left[1+ \sum\limits_{j=1}^\infty \left( 1- \left( 1 - \left(1-p\right)^j \right)^n  \right)
   \right]^{-1}.
\end{equation}

The derivation of Eq.~(\ref{eq:Result_no_multiplexing}) is provided in Appendix~\ref{sec:ProofTn}, but this result (in a different form) was already known in the context of chains of quantum repeaters between two participants~\cite{Bernardes2011}. 
To show this, let us simplify Eq. (\ref{eq:Result_no_multiplexing}):
\begin{align}
    \langle L_1 \rangle 
    &=
    \left[
    1+\sum_{j=1}^\infty
    \sum_{k=1}^n
    (-1)^{k+1}
    \binom{n}{k}
    (1-p)^{jk}
    \right]^{-1}
    \nonumber\\
    &=
    \left[
    1+\sum_{k=1}^n
    (-1)^{k+1}
    \binom{n}{k}
    \frac{(1-p)^k}{1-(1-p)^k}
    \right]^{-1}
    \nonumber\\
    &=
    \left[
    \sum_{k=1}^n
    \frac{(-1)^{k+1}}
    {1-(1-p)^k}
    \binom{n}{k}
    \right]^{-1}.
    \label{eq:l1alt}
\end{align}
The last expression in the square brackets is well-known in probability theory: It is the expectation value of the maximum of $n$ independent geometrically distributed random variables with the success probability $p$~\cite{Szpankowski1990}. In our context, this represents the average waiting time for the GHZ measurement when starting from empty memories. 
Indeed, the geometric distribution is the probability distribution of the number of Bernoulli trials (random experiments with exactly two possible outcomes) to get one success. 
For a successful GHZ measurement, all $n$ parties must have a filled memory, hence the waiting time is the maximum number of Bernoulli trials over all participants.  

Then, $\langle L_1\rangle$ is the inverse of the expectation value of the maximum of $n$ independent geometrically distributed random variables. This is due to the absence of multiplexing: Once a GHZ measurement is performed, the system returns to the initial state with all memories empty.

The derivation in Appendix~\ref{sec:ProofTn} can be considered as an alternative derivation of the expectation value of the maximum of $n$ independent geometrically distributed random variables. The advantage of Eq.~(\ref{eq:l1alt}) is that the denominator contains a finite sum, but the terms have alternating signs. The advantage of Eq.~(\ref{eq:Result_no_multiplexing}) is that all terms have the same sign.

The expression in the square brackets in Eq.~(\ref{eq:l1alt}) for the waiting time was obtained in a model of two parties connected by a sequence of $n$ segments with $n-1$ repeaters~\cite{Bernardes2011} (see also Refs.~\cite{Praxmeyer2013,Shchukin2019}), which is, as we noticed above, mathematically equivalent to the considered model of a multipartite quantum repeater.

Known estimates on the expectation value of the maximum of $n$ independent geometrically distributed random variables are~\cite{Eisenberg2008}:
\begin{equation}
\label{eq:WaitTimeEstimates}
    \frac1{\ln\frac1{1-p}}
    \sum_{k=1}^n\frac1k
    \leq
    \langle L_1\rangle^{-1}
    \leq
    1+\frac1{\ln\frac1{1-p}}
    \sum_{k=1}^n\frac1k.
\end{equation}
Recall that
\begin{equation}
    \sum_{k=1}^n\frac1k=\ln n+\gamma+\frac{1}{2n}+O(n^{-2}),
\end{equation}
where $\gamma=0.57721\ldots$ is the Euler-Mascheroni constant. Substitution of Ineqs.~(\ref{eq:WaitTimeEstimates}) into Eq.~(\ref{eq:l1alt}) leads to the bounds for the stationary repeater rate in the case of no multiplexing. Thus, the rate $\langle L_1\rangle$ decreases slowly as $(\ln n)^{-1}$ for a large number of communicating parties $n$. 

\par
\begin{figure}
    \centering
    \includegraphics[width=\linewidth]{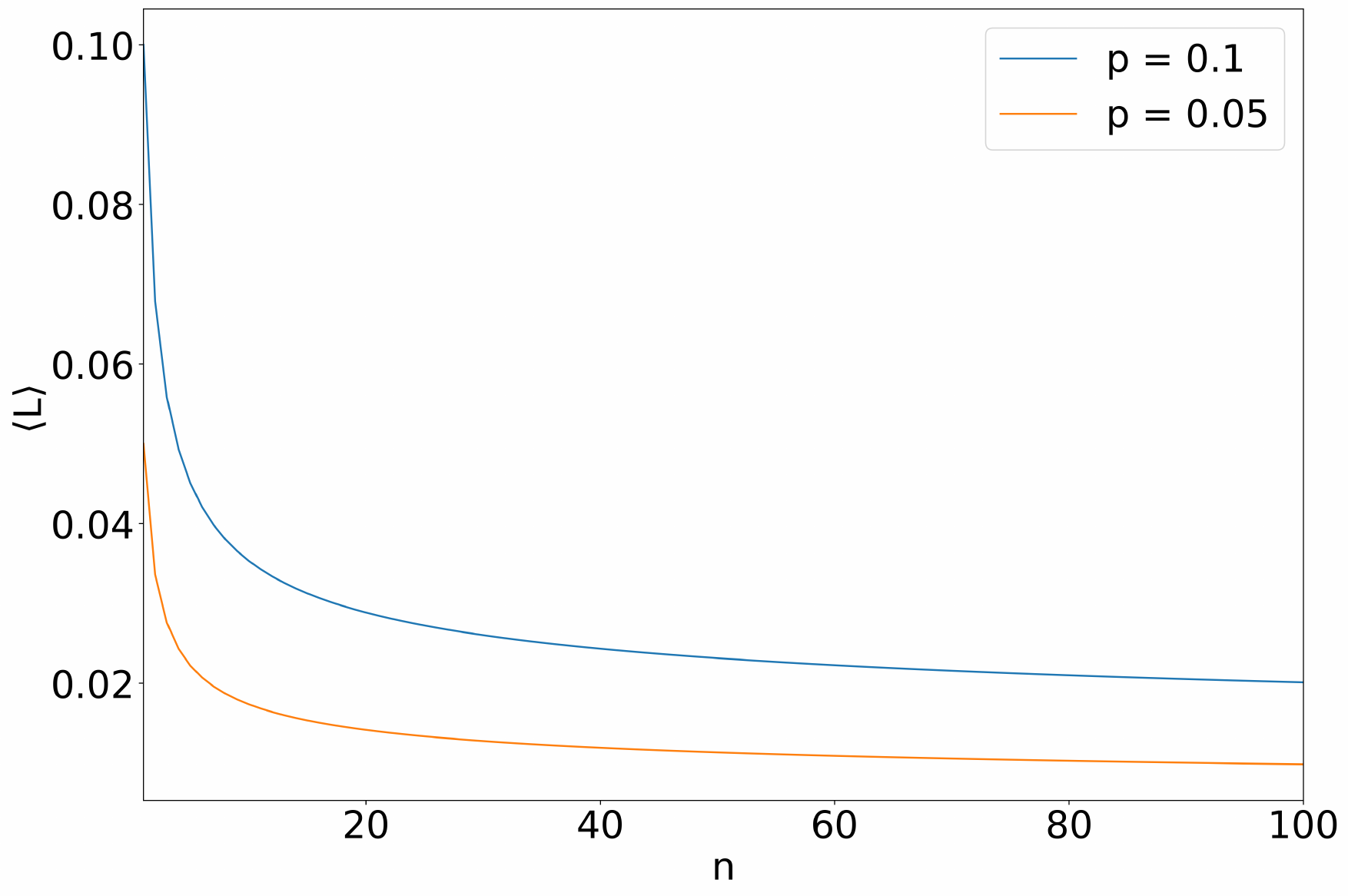}
    \caption{Multipartite quantum repeater without multiplexing: average number of GHZ measurements performed in the asymptotic limit following Eq.~(\ref{eq:Result_no_multiplexing}) for two different success probabilities $p=0.1$ and $p=0.05$, as a function of the number of parties $n$.} 
    \label{fig:Rate_no_multiplexing}
\end{figure}

Fig.~\ref{fig:Rate_no_multiplexing} shows the rates for two different success probabilities. 
One can see that, although the decrease in the average repeater rate is slow for a large number of communicating parties $n$, it decreases rapidly for small $n$. 
This motivates the use of multiple memories per party, allowing multiplexing to be integrated into the key distribution protocol~\cite{Multiplexing, kunzelmann_paper}, which is discussed in the following section.

\section{The case of  multiplexing}
\label{sec:Generalization}

\subsection{Reduction of the configuration space}
\label{sec:Reduction}

Now consider the case of memory multiplexing, i.e., $m\geq2$. We already mentioned that the number of GHZ measurements depends only on the numbers of the filled memories for each party $|\vec a_i|$ [see Eq.~(\ref{eq:ell})], but not on their positions. Hence, we can merge the configurations corresponding to the same tuples $(|\vec a_i|)_{i=1}^n$. We will denote $|\vec a_i|=k_i$ and $(k_1,\ldots,k_n)=\mathbf{k}$. Then the number of configurations is reduced from $2^{nm}$ to $(m+1)^n$: Each of the $n$ parties is fully characterized by the number of filled memories, from 0 to $m$. 

Then we can interpret $\pi$ as a vector of length $(m+1)^n$, and $\sigma$, $\mu$, and $T=\mu\sigma$ as $(m+1)^n\times(m+1)^n$ matrices. 

The storage map is given by the tensor power $\sigma = (\sigma^{(1)})^{\otimes n}$, where the single-particle storage map is given by 
\begin{equation}
       \sigma^{(1)}_{k'k} = 
       \begin{cases}
    0, & k > k', \\ 
    \binom{m-k}{k'-k}(1-p)^{(m-k')}p^{(k'-k)},&  \text{otherwise,} 
    \end{cases} 
\end{equation}
with $k,k' \in \{0, 1, \hdots, m \}$. The matrix of the measurement map is defined as follows: For arbitrary $\mathbf k=(k_1,\ldots,k_n)$ and $\mathbf k'=(k'_1,\ldots,k'_n)$, where $k_i,k'_i\in\{0,\ldots,m\}$,
\begin{equation}
    \mu_{\mathbf k'\mathbf k}
    =\begin{cases}
        1,&\mathbf{k}-\mathbf{k'}=l\cdot\vec{1}
        \text{ and }
        \min\mathbf k'=0,
        \\
        0,&\text{otherwise},
    \end{cases}
\end{equation}
where $\vec{1}=(1,\ldots,1)$ is the vector containing $n$ ones, $\min\mathbf{k'}$ is the minimum over $k'_1,\ldots,k'_n$, and $l\in\{0,\ldots,m\}$ is the number of GHZ measurements. Then Eq.~(\ref{eq:ell}) becomes
\begin{equation}
\label{eq:ellreduced}
    \langle L\rangle
    =\sum_{\mathbf k=\{0,\ldots,m\}^n}
    (\sigma\pi^*)_{\mathbf k}
    (\min \mathbf{k}).
\end{equation}

We can apply one more reduction of the configuration space if we merge configurations that differ only by permutations of the parties. 
That is, $\mathbf k=(k_1,\ldots,k_n)$ and $\mathbf k'=(k'_1,\ldots,k'_n)$ are considered equivalent if $(k'_1,\ldots,k'_n)$ can be obtained from $(k_1,\ldots,k_n)$ by a permutation of elements. Then, we can define a ``canonical'' order of $k_i$ as, e.g., decreasing: $k_1^\downarrow\geq k_2^\downarrow\geq\ldots\geq k_n^\downarrow$ and the state is described by an ordered $n$-tuple $(k_1^\downarrow,\ldots,k_n^\downarrow)$.

\subsection{Bipartite setup ($n=2$) for small success probabilities \textit{p}}

\begin{figure}
    \centering
    \includegraphics[scale=0.21]{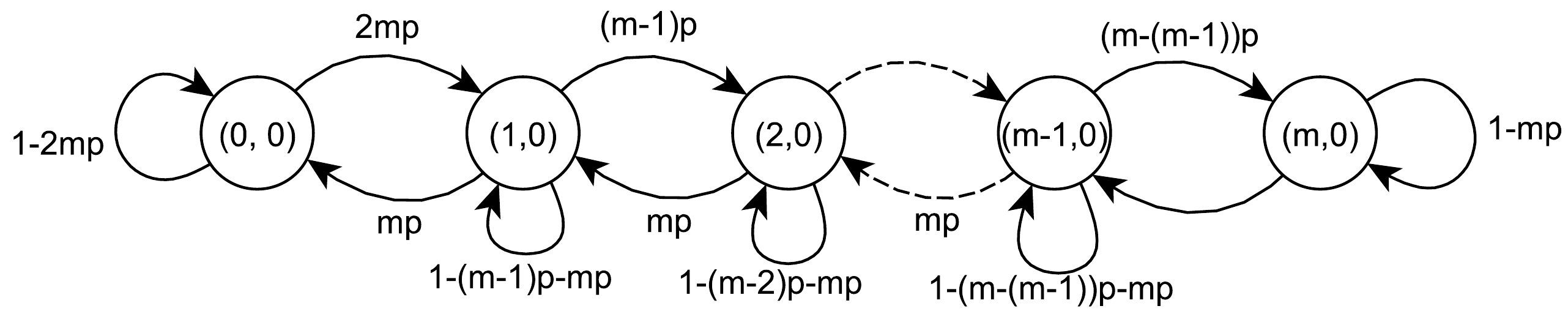}
    \caption{Graph representation for the bipartite ($n=2$) star graph assuming that the success probability $p$ of filling a memory cell is small so that it only appears in first order and at most one memory cell can be filled in each storage step. The vertices $(k_1^\downarrow,k_2^\downarrow)$ are occupation numbers of parties' memory cells in decreasing order. The last number is always zero because as soon as the corresponding party also fills one memory cell, a GHZ measurement is immediately performed, which empties this memory again. 
 }
    \label{fig:graph_bipartite}
\end{figure}

For the bipartite case (i.e., $n=2$) with multiplexing~\cite{Multiplexing}, a general expression can be deduced under the assumption of a small Bell pair generation probability $p \ll 1$. Then, all formulas are approximated by the first order in $p$, e.g., $(1-p)^m\approx1-mp$. Physically, that means that at most one memory can be filled per round. Nevertheless, qubits are kept in the memories over rounds once they are stored. The advantage is that the transition matrix has fewer entries that are unequal to 0. Consequently, the graph representing the Markov chain reduces significantly (see Fig.~\ref{fig:graph_bipartite}). 

We use the description of the configuration space from the end of Sec.~\ref{sec:Reduction}. Namely, a configuration is a pair $(k_1^\downarrow,k_2^\downarrow)$ of parties' occupation numbers in decreasing order.  
For example, the vertex $(1,0)$ includes all configurations where one party has one filled memory.  
Since we empty memories (by performing the GHZ measurement) at the end of a round, configurations where both parties have filled memories, such as $(1,1)$, cannot be reached after each iteration, as it ends with a measurement step. Therefore, these configurations do not appear in the graph shown in Fig.~\ref{fig:graph_bipartite}.
Including all assumptions, the number of vertices, each representing one configuration, reduces to $m+1$ for the bipartite setup. 

It is worthwhile to note that the transition probability from $(k_1^\downarrow,k_2^\downarrow)=(0,0)$ to $(k_1^\downarrow,k_2^\downarrow)=(1,0)$ is $2mp$ rather than $mp$ because any of the two parties can fill one of its memories. In other words, the ordered configuration $(k_1^\downarrow,k_2^\downarrow)=(1,0)$ corresponds to the two unordered configurations $(k_1,k_2)=(1,0)$ and $(k_1,k_2)=(0,1)$, and the transition probability from $(0,0)$ to each of these configurations is $mp$.  However, the transition probability from $(k_1^\downarrow,k_2^\downarrow)=(1,0)$ to $(k_1^\downarrow,k_2^\downarrow)=(2,0)$ is $(m-1)p$, because now the order of the parties is fixed.
\\
\par
Recall that a Markov chain can be represented as a directed graph, where the vertices $V$ are configurations and the edges $E$ are possible transitions between them in one time step (with nonzero probabilities). In our case, $V=\{0,\ldots,m\}^n$. Then, the stationary distribution in the asymptotic limit can be obtained via the Markov chain tree theorem:
\begin{theorem}[\cite{MCTree}]
    Let a stochastic matrix $T$ define an irreducible finite Markov chain with stationary distribution $\pi^*=(\pi^*)_{\mathbf{k}\in V}$. Then
    \begin{align}
        \pi^*_{\mathbf k} = \frac{\| \mathcal{A}_{\mathbf k}\|}{\| \mathcal{A}\|}.
    \end{align}
\end{theorem}

Here, $\mathcal{A}_{\mathbf k}$ denotes the set of arborescences of a chosen root $\mathbf k$, and $\mathcal{A}$ is the set of all arborescences of the graph $G$. 
 An arborescence with root $\mathbf k$ is a set of edges $A\subseteq E$, that fulfills the following properties:
 \begin{itemize}
     \item each vertex $\mathbf k' \in {V}$ has a directed path in the subgraph $G' = ({V}, A)$ to the chosen root $\mathbf k$.
     \item each vertex, except the root $\mathbf k$, has precisely one outgoing edge. 
     \item the root $\mathbf k$ has no outgoing edge. 
 \end{itemize}
 Each $\pi_{\mathbf k}$ is the probability of finding configuration $\mathbf k$ in the multipartite quantum repeater. In the graphical representation, each vertex $\mathbf k$ represents one such configuration. 
 The weight $\|A\|$ of an arborescence $A$ is given by the product of the weights of the edges $p_e$ (in our case, probabilities of the corresponding transitions) included in that arborescence:
 \begin{align}
 \label{eq:MatTreeTh}
    \|A\| = \prod_{e \in A} p_e.
 \end{align}
Then, $\| \mathcal{A}_{\mathbf k}\|$ is the sum over the weights of all arborescences within the set $ \mathcal{A}_{\mathbf k}$ and $\| \mathcal{A}\|$ is the sum over the weights of all arborescences of the graph $G$:
\begin{equation}
\begin{split}
    \|\mathcal{A}_{\mathbf k}\|&=\sum_{A\in\mathcal{A}_{\mathbf k}}
    \|A\|,
    \\
    \|\mathcal{A}\|&=\sum_{A\in\mathcal{A}}
    \|A\|
    =
    \sum_{\mathbf k\in V}
    \|\mathcal{A}_{\mathbf k}\|.
\end{split}
\end{equation}
Thus, 
to calculate $\pi'_{\mathbf k}$ for all roots ${\mathbf k}$ (i.e., all configurations), it is required to find all arborescences in the graph and calculate their weights. 
The Markov chain tree theorem can also be applied to graphs with self-loops, which, however, do not contribute to arborescences and can thus be ignored~\cite{MCTree}.

\begin{figure}
    \centering
    \includegraphics[scale=0.2]{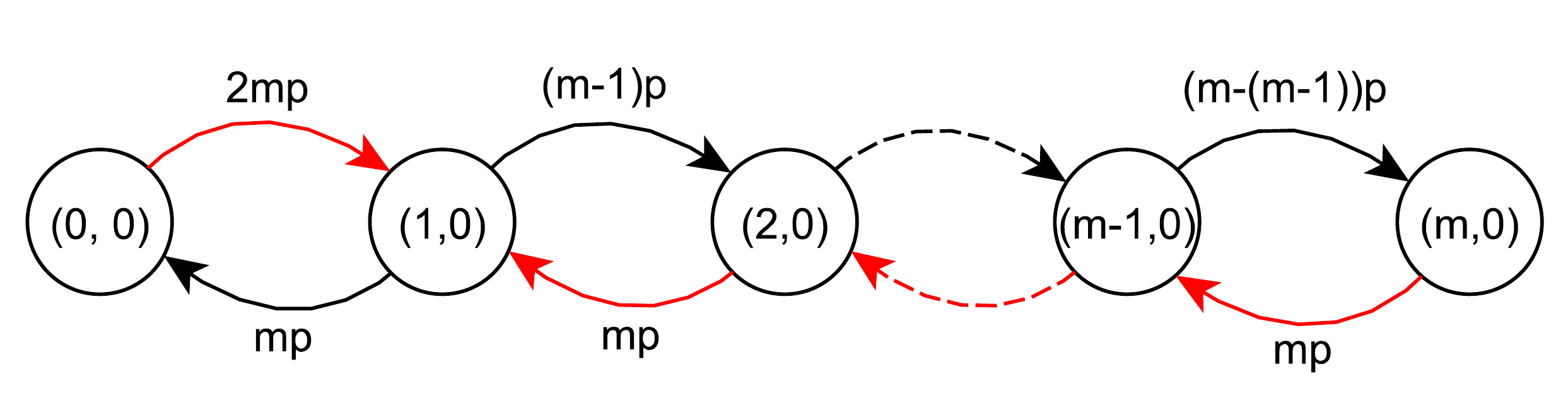}
    \caption{The only arborescence $A_{(1,0)}$ for the second node (root) $(1,0)$, i.e., the set of edges that lead to a directed path from all nodes to the chosen root $(1,0)$, is shown in red. Since self-loops do not contribute to the calculation, they are not shown in the graph representation here.  }
    \label{fig:example_Baum}
\end{figure}

We find that, in our case, each root has a single arborescence only:
Starting from each end of the chain, all edges pointing towards the root are part of the arborescence. 
An example for the arborescence $A_{(1,0)}$ of root $(1,0)$ is shown in Fig.~\ref{fig:example_Baum}.  
By applying the Markov chain tree theorem to the bipartite setup, 
we find for the weights $W_{\mathbf k}$ for each root ${\mathbf k} \in \{ (0,0), (1,0), \hdots, (m,0) \}$ the following expressions:
\begin{align}
    \|\mathcal A_{(0,0)}\| &= (mp)^m,  \\
    \|\mathcal A_{(1,0)}\| &= 2(mp)^m, \\
    \|\mathcal A_{(k,0)}\| &= 2m^{m-k+1} p^m \frac{(m-1)!}{(m-k)!} \text{  \hspace{0.1cm}  for \hspace{0.1cm}} 1 \leq k \leq m.
\end{align}
To calculate the average number of GHZ measurements $\langle L \rangle$ per round, we use Eq. (\ref{eq:ellreduced}). Due to the assumption that at most one memory can be filled per round, maximally one GHZ measurement can be performed.  
The stationary distribution is calculated with Eq. (\ref{eq:T}). 
We see from the graph that only the configuration $\mathbf k=(0,0)$ at the beginning of a round leads to no GHZ measurement since, in our approximation, only one memory can be filled in one round. All other configurations $\mathbf k \in \{ (1,0), (2,0), \hdots, (m,0) \}$ lead to a GHZ measurement with the probability $p m$ of increasing $k_1$ from 0 to 1.  For the average number of GHZ measurements, we then find: 
\begin{equation}
\begin{split}
\label{eq:binary_rate}
    \langle  L \rangle 
    &= p m \left(\frac{\sum_{k'=1}^m \|\mathcal A_{(k',0)}\|}{\sum_{k=0}^m \|\mathcal A_{(k,0)}\|} \right)    \\
    &= p m  \left(1 - \frac{\|\mathcal A_{(0,0)}\|}{\sum_{k=0}^m \|\mathcal A_{(k,0)}\| } \right)   \\
    &= p m  \left( 1 - \frac{1}{1 + 2\sum_{k=1}^m m^{-k+1} \frac{(m-1)!}{(m-k)!}}   \right).  
\end{split}
\end{equation}
For the repeater rate, it then holds:
\begin{align}
\label{eq:RR}
    R_\infty
    &= p  \left( 1 - \frac{1}{1 + 2\sum_{k=1}^m m^{-k+1} \frac{(m-1)!}{(m-k)!}}   \right)   
\end{align}
\begin{figure}
    \centering
    \includegraphics[width=\linewidth]{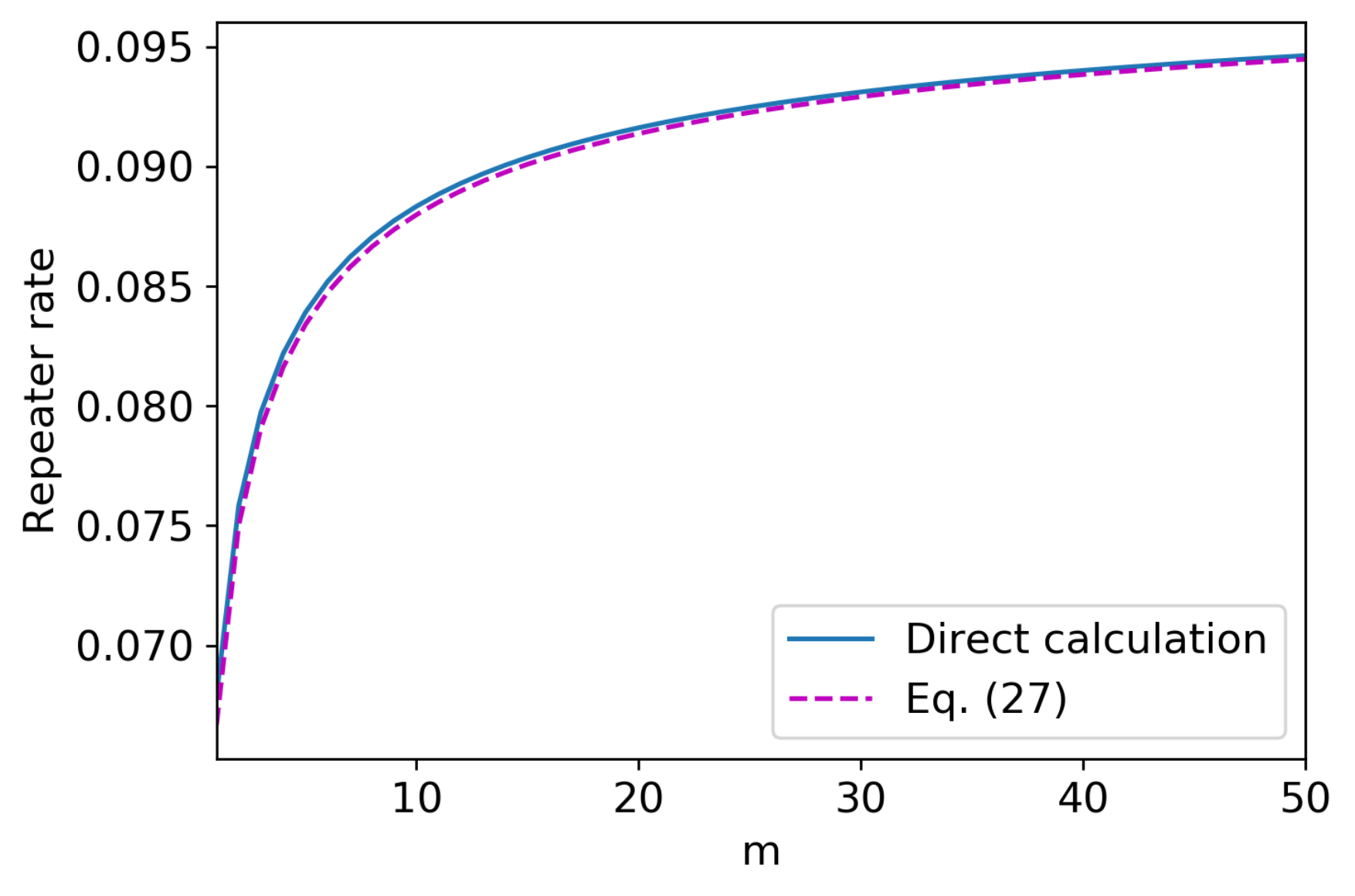}
    \caption{Repeater rate $R_\infty=\langle L\rangle/m$ for a bipartite network for different number of memories per party $m$. The success probability is chosen to be $p=0.1$. The analytical approximation is calculated via Eq. (\ref{eq:RR}), while the direct calculation is obtained via $T^n \pi_{init}$ for large $n$. }
    \label{fig:BipartiteRouterRate}
\end{figure}
We can find asymptotic behavior for large $m$ as follows: 
\begin{multline}
    \sum_{k=1}^m m^{-k+1} \frac{(m-1)!}{(m-k)!}=
    \frac{m!}{m^m}
    \sum_{k=0}^{m-1}\frac{m^{k}}{k!}
    \\
    \approx
    \sqrt{2\pi m} \,e^{-m}
    \sum_{k=0}^{m-1}\frac{m^{k}}{k!}=\sqrt{2\pi m}\,F_m(m-1),
\end{multline}
where $F_m$ is the cumulative distribution function of the Poisson probability distribution with the expectation $m$. Due to the central limit theorem, the Poisson distribution is approximated by the normal distribution for large $m$, which is symmetric with respect to the expectation value. Hence, $F_m(m-1)\approx1/2$ for large $m$ (actually, already for $m=5$ the approximation is good). We obtain then
\begin{equation}
\label{eq:largemsimp}
    R_\infty
    \approx p  \left(1 - 
    \frac{1}{1+\sqrt{2\pi m}}   \right).
\end{equation}

In Fig.~\ref{fig:BipartiteRouterRate}, we compare the repeater rate for the bipartite network with up to $m=50$ memories per party and a success probability of $p = 0.1$ determined in two different ways. 
We compare it with a direct calculation of $T^n \pi_{init}$ for large $n$.  
In the approximate solution, we use Eq. (\ref{eq:RR}) to calculate the repeater rate. Fig.~\ref{fig:BipartiteRouterRate} shows that the approach of small $p$ leads to a lower bound that provides good results also for larger $m$.

\subsection{Larger network sizes}
\label{sec:largernetworks}

\begin{figure}
    \centering
    \includegraphics[width=\linewidth]{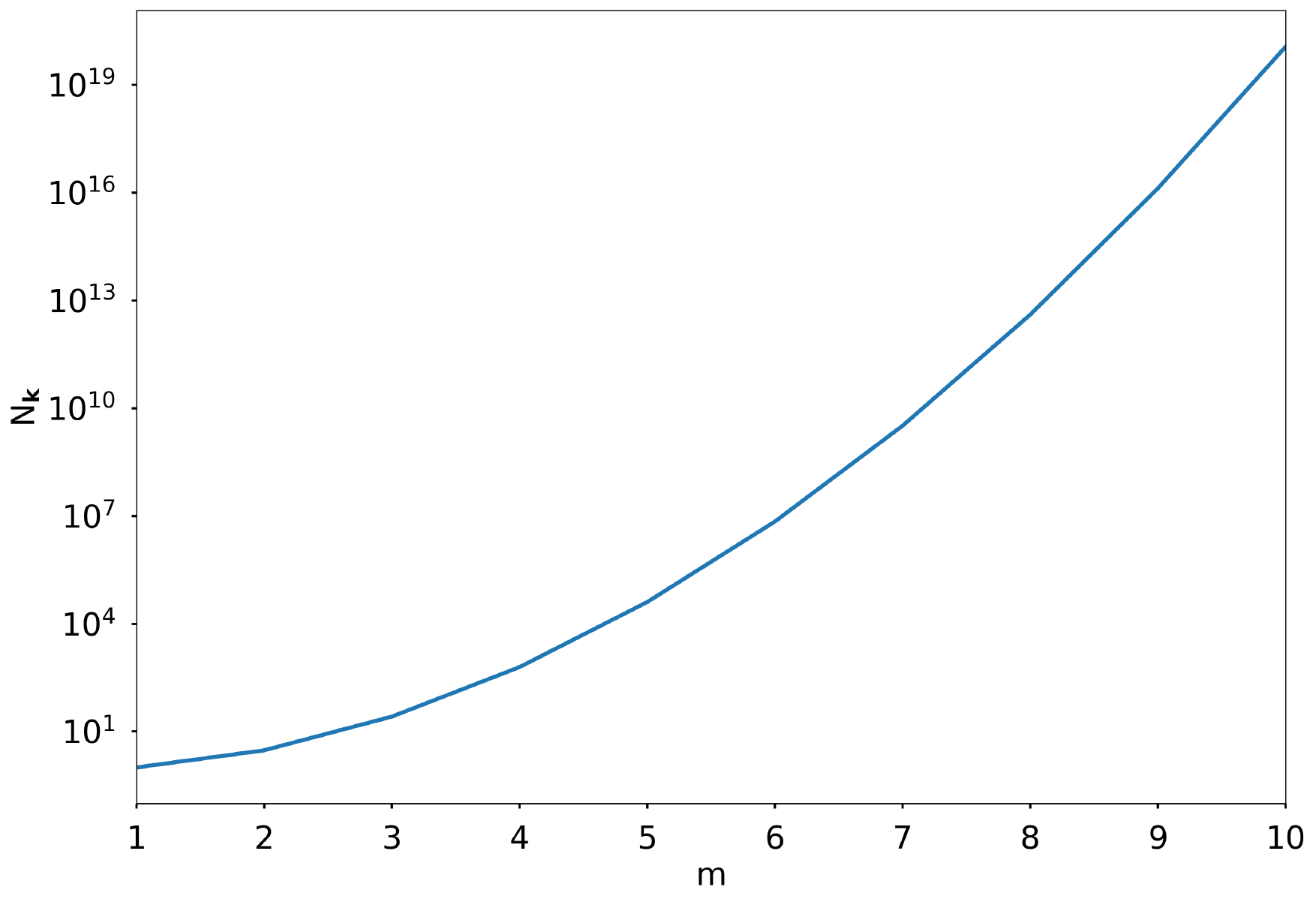}
    \caption{Number of arborescences $N_\mathbf{k}$ for root $\mathbf{k}=(m,m, 0)$ for a tripartite network with up to ten memories. Tutte's directed matrix-tree theorem (i.e., Eq.(~\ref{eq:Idontknow})) is used for the calculation. }
    \label{fig:Overview_Nj}
\end{figure}

For networks with $n>2$ parties and memory multiplexing $m\geq1$, it is hard to find a generalization by proceeding analogously to the bipartite network, even under the assumption that $p \ll 1$. This is because the resulting graphs representing the Markov chains have more than one arborescence per root $i$. This holds already in the simplest case with $n=3$ and $m=2$. Here, each root has three arborescences. 
The number of arborescences can be determined using Tutte's directed matrix-tree theorem:
\begin{theorem}[\cite{Tutte}]
    Let $G=(V,E)$ be a directed graph and $\mathcal{L}$ a matrix with entries
    \begin{equation}
     L_{\mathbf{k},\mathbf{k}'} =
  \begin{cases}
   \deg_{in} (\mathbf{k}'),        & \text{if } \mathbf{k}' = \mathbf{k}, \\
   -1,        & \text{if } \mathbf{k} \neq \mathbf{k}' \text{ and } (\mathbf{k}, \mathbf{k}') \in E, \\
   0 & \text{otherwise,}
  \end{cases}
\end{equation}
$\mathbf{k}, \mathbf{k}'\in V$, where $\deg_{in}(\mathbf{k}')$ is the in-degree of vertex $\mathbf{k}'$.
The number of arborescences $N_\mathbf{k}$ with root $\mathbf{k}\in V$ is then given by 
\begin{align}
\label{eq:Idontknow}
N_\mathbf{k} = \det \left( \Hat{\mathcal{L}}_\mathbf{k} \right) 
\end{align}
where $\hat{\mathcal{L}_{\mathbf{k}}}$ is the matrix produced by deleting the $\mathbf{k}$-th row and column from $\mathcal{L}$. 
\end{theorem}
Fig.~\ref{fig:Overview_Nj} shows the increase of $N_\mathbf{k}$ up to $m=10$ memories for a tripartite network for the first reduction of the configuration space described in Sec.~\ref{sec:Reduction}, i.e., the configuration is given by the (unordered) occupation numbers of three parties $(k_1,k_2,k_3)$, under the approximation $p\ll1$ (again, $p$ is taken into account only up to the first order).  
It can be seen that even under the assumption of small $p$, it is not possible to find a generalized expression for $\langle  L \rangle$. 
\\
\\
\par
 Nevertheless, we can construct bounds for the average number of GHZ measurements for the case of $n$ parties each having $m$ memories. Suppose we treat the memories independently of each other (i.e., no multiplexing is performed despite the multimemory setup). 
 This case leads to the following lower bound:
\begin{align}
\label{eq:LowBndSimp}
    \langle  L \rangle \geq m  \langle  L_1 \rangle
\end{align}

An upper bound for $\langle  L \rangle$ can be achieved from numerical simulations performed analogously to Ref.~\cite{kunzelmann_paper}: 

\begin{equation}
\label{eq:UpBndSimp}
    \langle  L \rangle \leq p m. 
\end{equation}
One can understand it as follows. After the measurement step, at least one party has no filled memories. The average number of memories filled in the next storage step for this party is $pm$. Since $L$ cannot be larger than the minimal number of filled memories among the parties, $\langle L\rangle$ cannot be larger than $pm$.

However, comparison with simulation show that both bounds (\ref{eq:LowBndSimp}) and (\ref{eq:UpBndSimp}) are loose. In the following subsection, we derive an approximate formula for the general repeater rate, making different approximations.

\subsection{Probabilistic model of the multipartite quantum repeater and the repeater rate for the general case}

The purpose of this subsection is to analyze the general case of an arbitrary number of parties $n$ and the number of memories per party $m$. Up to now, we have rigorous bounds only for the cases $m=1$ and arbitrary $n$ (Eqs.~(\ref{eq:Result_no_multiplexing}) and (\ref{eq:l1alt})) and $n=2$  and arbitrary $m$ (Eq.~(\ref{eq:binary_rate})). As we saw above, it is hard to generalize Eq.~(\ref{eq:binary_rate}) to the case of an arbitrary $n$. 
We need a formula from which the asymptotic behavior with respect to $n$ and $m$ could be easily derived.

Our derivation of such a formula will be based on the following probabilistic model of the multipartite quantum repeater:

\begin{itemize}
    \item 
     $X_{i,k}$ is the random variable of the number of filled memories (from 0 to $m$) for the $i$th party ($i=1,\ldots,n$) after the round $k$, i.e., before the $(k+1)$th storage stage, $t=0,1,2,\ldots$ The initial number of filled memories is $X_{i,0}=0$.

     \item $Y_{i,k}$ is the number of new filled memories for the party $i$ in the storage stage in the round $k$. It takes values from 0 to $m-X_{i,k}$ and obeys the binomial distribution:
     \begin{equation}
     \label{eq:PrYX}
         \Pr[Y_{i,k}=y|X_{i,k-1}=x]=\binom{m-x}{y}p^y(1-p)^{m-x-y},
     \end{equation}
     $y=0,1,\ldots,m-x$.

     \item $Z_{i,k}=X_{i,k-1}+Y_{i,k}$ is the number of the filled memories for the party $i$ before the measurement in the round $k$.

     \item $L_k=\min_{i}Z_{i,k}$ is the number of GHZ measurements in the round $k$.

     \item Measurement stage:
     \begin{equation}
     \label{eq:XZL}
         X_{i,k}=Z_{i,k}-L_k.
     \end{equation}

     \item $\langle L\rangle=\lim_{k\to\infty}\langle L_k\rangle$
     is the long-term expectation value of $L_k$, which we want to find (or approximate). If the probability distribution of $L_k$ converges to a limiting distribution, $L$ can be considered a random variable with this distribution.
\end{itemize}

However, the probabilistic analysis of the model described above is problematic due to dependencies of $X_{i,k}$ on each other by means of the subtraction of $L_k$, which depend on $X_{i,k-1}$ for all $i$. To break this dependency, we develop a simplified model. To do this, let us iterate Eq.~(\ref{eq:XZL}):
\begin{equation}
    X_{i,k}=X_{i,0}+
    \sum_{k'=1}^k Y_{i,k'}-
    \sum_{k'=1}^k L_{k'}.
\end{equation}
The intuition related to the law of large numbers tells that, for large $k$, $\sum_{k'=1}^k L_{k'}$ can be replaced by $k\langle Y\rangle$. In other words, random $L_{k'}$ can be replaced by a fixed number. 

In view of this intuition, consider the random variables $\tilde X_{i,k}$, $\tilde Y_{i,k}$, $\tilde Z_{i,k}$, and $\tilde L_k$ with the following modification: the recurrence equation (\ref{eq:XZL}) is replaced by
\begin{equation}
    \label{eq:XZl}
    \tilde X_{i,k}=\tilde Z_{i,k}-l
\end{equation}
for a \textit{fixed} number $l$ to be determined. Subsequently, the random variables $\tilde X_{i,k}$ can take negative values. Namely, if $\tilde X_{i,0}=0$, then $X_{i,k}$ can take arbitrary integer values from $-lk$ (in the case $Y_{i,k'}=0$, $k=1,\ldots,k$) to $m$. Then, $Y_{i,k}$ also can take values larger than $m$ (if $X_{i,k-1}$ is negative). Formula (\ref{eq:PrYX}) holds with $X$ and $Y$ replaced by $\tilde X$ and $\tilde Y$.

In this case, $X_{i,k}$ and $X_{j,k}$ are independent for $i\neq j$ (and identically distributed because they obey the same recurrence relation (\ref{eq:XZl})). In other words, the dynamics of the memory occupation numbers for different participants are uncoupled, which largely simplifies the analysis.

In the original model all $X_{i,k}$ are nonnegative or, equivalently,
\begin{equation}
    \min(X_{1,k},\ldots,X_{n,k})\geq0.
\end{equation}
In the simplified model, we replace this requirement with the following one: The number $l$ is chosen as a maximal value such that the inequality
\begin{equation}
\label{eq:EminPositive}
    \lim_{k\to\infty}
    \langle
    \min(\tilde X_{1,k},\ldots,\tilde X_{n,k})
    \rangle
    \geq0
\end{equation}
holds,
i.e., the long-time average of the minimum of the occupation numbers is non-negative. The intuition behind such a replacement is again the law of large numbers: we can hope that, in the asymptotic case, the fluctuations around the expectation value are negligible.

One can suggest the following financial analogy. If $\tilde X_{i,k}$ is the current ``wealth'' of the $i$th party, we take a fixed ``tax'' $l$ and allow the party to ``borrow'' money for some time (negative $\tilde X_{i,k}$), but, on average, the minimal ``wealth'' among the parties must remain non-negative.

Then, under the additional assumption of normal distribution for $X_{i,k}$ and in the asymptotic case of large $n$, one can derive the following formula for the maximal $l$, which is an estimate for $\langle L\rangle$ in the original model (see Appendix~\ref{sec:ProofSimpModel}):

\begin{equation}
\label{eq:ellgen}
    \langle L \rangle
    \approx
    p m
    \left(
    \sqrt{
    \frac{\alpha^2\beta}{4m}
    \ln n
    +1
    }
    -
    \sqrt{
    \frac{\alpha^2\beta}{4m}
    \ln n
    }
    \right)^2.
\end{equation}
Here $\beta=(1-p)/(2-p)$,
\begin{equation}
    \alpha=
    \sqrt{2}
    \left(
    1-
    \frac{\ln(4\pi\ln n)-2\gamma}
    {4\ln n}
    \right)
\end{equation}
and $\gamma$ is the Euler-Mascheroni constant. 

Note that for $n\to\infty$, we get 
\begin{equation}
\label{eq:LlargeN}
    \langle L\rangle
    \approx
    \frac{p m^2}
    {\alpha^2\beta\ln n},
\end{equation}
i.e., logarithmical decrease, which agrees with the results for the case of no multiplexing (Sec.~\ref{sec:nomulti}). The quadratic dependence on $m$ does not contradict the upper bound (\ref{eq:UpBndSimp}) with the linear dependence because, for large $n$, the right-hand side of Eq.~(\ref{eq:LlargeN}) is smaller than $pm$. The quadratic increase with $m$ can be understood by the double role of memory multiplexing. First, it increases the average number of new Bell links per round. E.g., if all memories of a party are empty, then $pm$ is the average number of filled memories after a storage step. Second, if a party already has one filled memory and waits for other parties, multiple memories allows him not to waste time, but to establish additional Bell links for future measurements, so, after the measurement step, the parties start not from scratch, but already have Bell links.

Interestingly, though strictly speaking, formula (\ref{eq:ellgen}) uses the approximation of large $n$, substitution of $n=1$ in Eq.~(\ref{eq:ellgen}) gives the exact result for this case $\langle L \rangle=p m$. 

Also we see that, for large $m$, the repeater rate (\ref{eq:Router_rate}) saturates:
\begin{equation}
\label{eq:RouterRateSaturation}    R_\infty=\frac{\langle L \rangle}m
    \approx p
    \left(1-
    \sqrt{
    \frac{\alpha^2\beta}{4m}
    \ln n
    }
    \right)
    ,
\end{equation}
so the effect of an additional memory for each party does not increase $R_\infty$. Also we see the agreement with the large $m$ asymptotics (\ref{eq:largemsimp}) for the case $n=2$: The repeater rate saturates as $c/\sqrt m$ with some constant $c$. Eqs.~(\ref{eq:largemsimp}) and (\ref{eq:RouterRateSaturation}) (for small $p$) give close values of $c$.
\\
\par
\begin{figure}
    \centering
    \includegraphics[width=\linewidth]{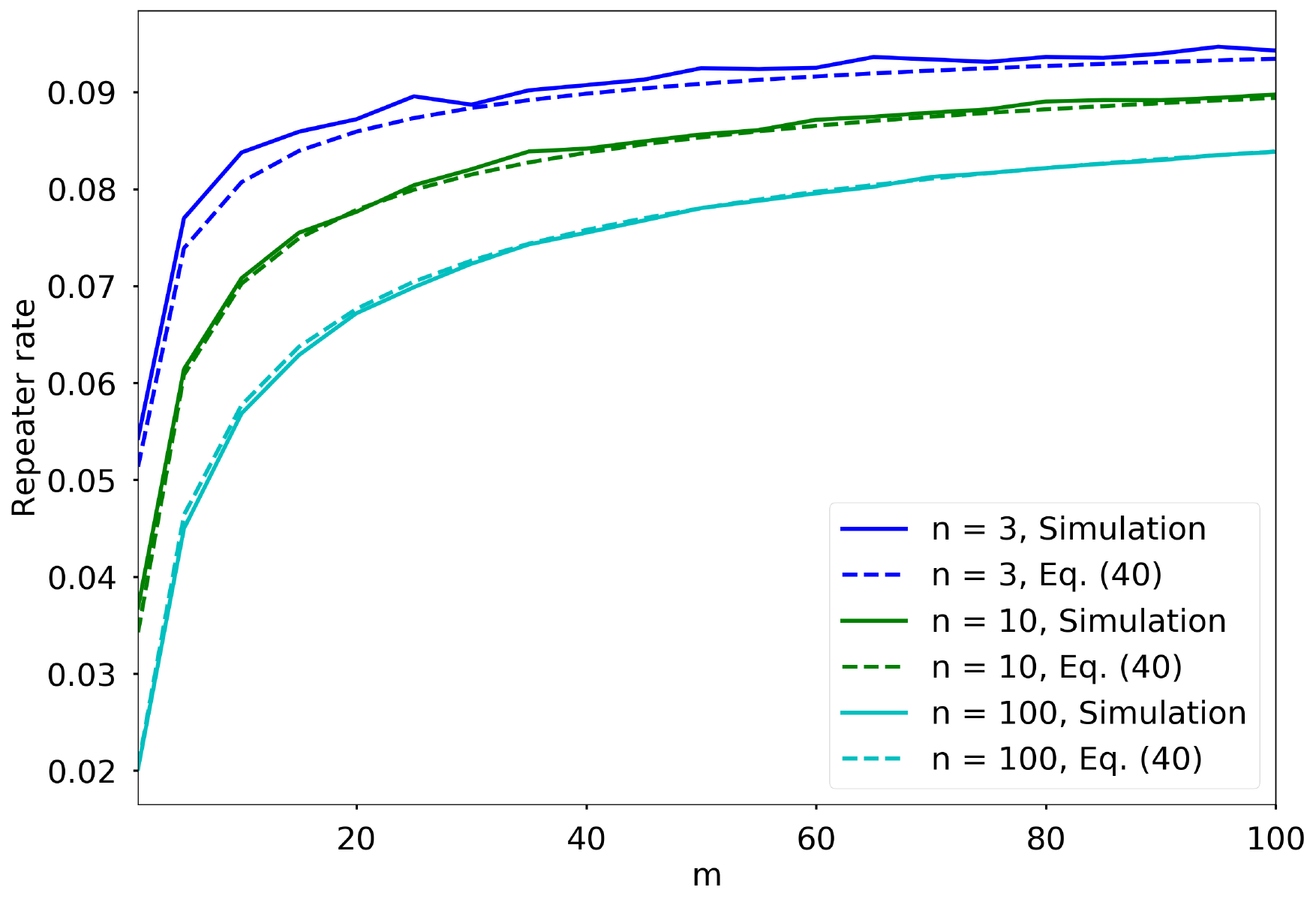}
    \caption{Comparison between simulations and the approximate Eq. (\ref{eq:ellgen}) for the repeater rate $R_\infty=\langle L \rangle/m$ with the probability of successfully storing entangled qubits in memory is $p=0.1$. The repeater rate is shown for varying numbers of parties $n$ and memories $m$. }
    \label{fig:Comparison}
\end{figure}

To justify the approximation, we show the repeater rate for varying network sizes in Fig.~\ref{fig:Comparison} comparing the approximation formula obtained in Eq.~(\ref{eq:ellgen}) and the simulation of the multipartite quantum repeater performed. 
Eq.~(\ref{eq:ellgen}) gives a very good approximation for the whole range of parameters (though formally it was derived for the asymptotic case of large $n$).
\\ \par
\begin{figure}
    \centering
    \includegraphics[width=\linewidth]{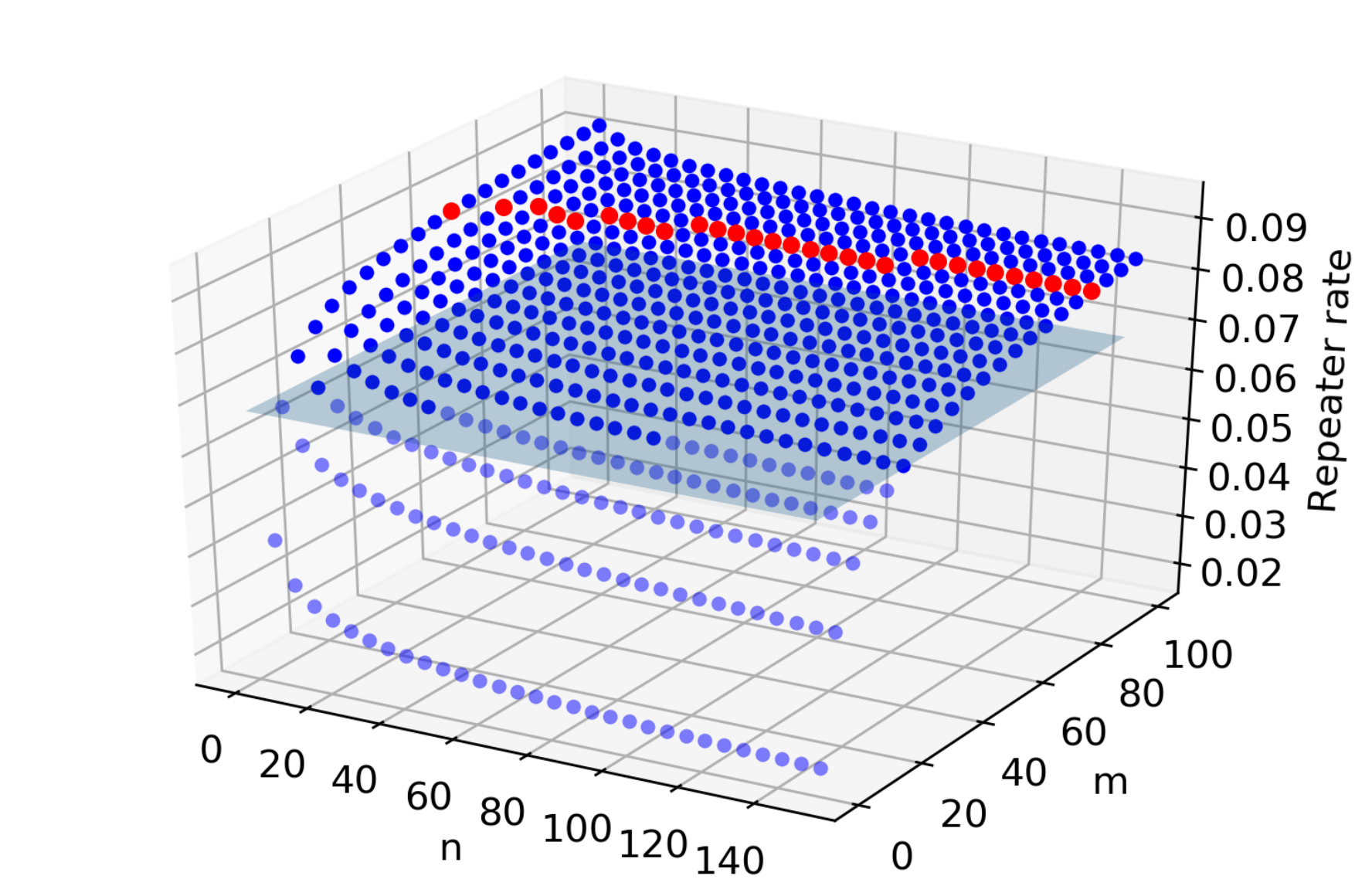}
    \caption{Repeater rate $R_\infty = \langle L \rangle/m$ obtained by formula (\ref{eq:ellgen}) for different network sizes depending on $n$ and $m$ and a success probability of $p=0.1$. 
    For a rate beyond the line of red dots, the rate does not further increase with the number of memories.
    The light blue plane shows the minimal achievable repeater rate in a bipartite setup with a single link for both parties. }
    \label{fig:scaling}
\end{figure}

In Fig.~\ref{fig:scaling}, we give an overview of the scaling of the repeater rate calculated via the approximation formula up to a network size of $n = 150$ parties, each having up to $m = 100$ memories.  
The plot shows that, for small $m$, the repeater rate grows linearly with $m$, but then the growth slows down. The plot confirms the saturation of the repeater rate (\ref{eq:RouterRateSaturation}). 
The red dots indicate the limit at which adding more memories to the network no longer significantly increases the repeater rate, i.e., the difference in the repeater rate becomes smaller than 0.0001 with increasing $m$. 
For networks with more than 100 parties, this limit is at about $m = 85$ per party. 
\par 
The light blue plane shown in Fig.~\ref{fig:scaling} gives the repeater rate achieved in a bipartite network with only one memory per party $(R_\infty = 0.068)$.  For all network sizes considered here, this threshold can be reached by increasing the number of memories per party. Thus, one can achieve the repeater rate of a bipartite network also in a multipartite network.

\section{Conclusion} 
\label{sec:summary}

Establishing entangled states between end nodes in large quantum networks is one of the main challenges to allowing the end nodes to communicate or share secret keys. To enlarge the distance between the end nodes and overcome distance-based fiber losses, quantum repeaters are used.  
In our work, we have analyzed the repeater rate of such multipartite quantum repeaters (multipartite quantum repeaters), i.e., the average number of distributed GHZ states per round in the stationary (long-term) regime.
We have considered both single links between the end nodes and the central multipartite quantum repeater and multiple links (multiplexing). Our results can be used to plan quantum networks and estimate the achievable repeater rates for various network sizes. The optimal number of memories can be calculated, especially when the number of parties increases.
\\ \par
For the single link case, we have derived  Eqs.~(\ref{eq:Result_no_multiplexing}) and (\ref{eq:l1alt}), which coincide with the rate of a chain of repeaters between two participants obtained in Refs.~\cite{Bernardes2011, Shchukin2019}. 
For the multiplexing setup and two parties, we have derived an approximate formula (\ref{eq:RR}) for the repeater rate based on the approximation of small success probabilities $p$.   

Finally, we have derived an approximate formula to calculate the average number of GHZ measurements per round for an arbitrary number of participants and memories.
This approximation gives very good agreement with the simulation. 
 
It turns out that the repeater rate saturates for a large number of memories, i.e., the number of GHZ measurements per round grows not faster than linearly with the number of memories per party. 
Additionally, we observe that as the number of participants $n$ increases, the repeater rate decreases slowly as $(\ln n)^{-1}$, provided the number of memories per party remains constant. 
\\
\par
In our work, we show that there is a maximal number of memories that helps to increase the repeater rate. Adding more memories does not lead to higher rates. In future work, integrating entanglement purification will be of great interest~\cite{Bennett_1996}. It will be interesting to analyze how the fidelities change due to the underlying network structure and how entanglement purification can increase the fidelity~\cite{Davies_2024}. It should be investigated for which network sizes multiplexing leads to higher fidelities than entanglement purification. In a further step, applications such as conference key agreement~\cite{CKA} should also be included in the analysis.

\section{Acknowledgment}

We thank Jörg Rothe and Moritz Wiese for valuable discussions.
This work was funded by the Federal Ministry of Education and Research BMBF (QR.X with Project No. 16K1SQ023 and QuKuK with Project No. 16KIS1618K).

\bibliography{bibliography.bib}

\appendix

\section{Derivation and proof of the asymptotic repeater rate without multiplexing}
\label{sec:ProofTn}

Here, we provide our derivation of the repeater rate for the setup with a single memory per party only. We start with the transition matrix $T'$ which is given by 
\begin{align}
    T' = \sigma  \mu = \left(\sigma^{(1)}\right)^{\otimes N}  \mu  \nonumber 
\end{align}
Note that for this derivation, we consider the transition matrix in reverse order. This can be done since the stationary distribution is independent of the initial distribution $\pi^{(0)}$. This transition matrix already provides the distribution after the process of storing qubits in memory. It is, therefore, not necessary to apply another map. This order is chosen here because it simplifies the following calculations. 
\par
The storage process for one party with a single memory is given by
\begin{align}
        \sigma^{(1)} =   
    \begin{pmatrix}
    1-p & 0 \\
    p & 1
  \end{pmatrix}
    \end{align}
with $p$ being the success probability of the optical fiber. 
The total storage map for $n$ parties follows from the tensor product:
\begin{align}
        \sigma =   
    \begin{pmatrix}
    1-p & 0 \\
    p & 1
  \end{pmatrix} ^{\otimes n}
    \end{align}
A measurement is performed only when the memory of every party is filled. In that case, the memories are emptied; in all other cases, the memory configuration does not change. Therefore, the measurement map is given by
\begin{align}
        \mu = \mathds{1}_{2^n} - 
        \begin{pmatrix}
    0 & 0 \\
    0 & 1
  \end{pmatrix}^{\otimes n} + 
  \begin{pmatrix}
    0 & 1 \\
    0 & 0
  \end{pmatrix}^{\otimes n}
    \end{align}
By combining both maps, we find the following transition map: 
\begin{align}
    T' 
    &= \sigma + \sigma {\left( - \begin{pmatrix}
    0 & 0 \\
    0 & 1
  \end{pmatrix}^{\otimes n} + 
  \begin{pmatrix}
    0 & 1 \\
    0 & 0
  \end{pmatrix}^{\otimes n} \right)} \nonumber \\
    &\equiv \sigma + \sigma  X_n
\end{align}
where we set ${\left( - \begin{pmatrix}
    0 & 0 \\
    0 & 1
  \end{pmatrix}^{\otimes n} + 
  \begin{pmatrix}
    0 & 1 \\
    0 & 0
  \end{pmatrix}^{\otimes n} \right)} = X_n$.
In the next step, we calculate $ T'^s$ to get the stationary distribution of the Markov chain $\pi^* = \lim_{ s\rightarrow \infty} T'^s \pi^{(0)} $. 
It is possible to rewrite $T'^s$ by 
\begin{align}
\label{eq:Tn}
    T'^s &= \sigma^s + \sum_{i = 1}^{s} A_i  X_n  B_{s-i}
\end{align}
with the second term given by
\begin{align}
\label{eq:Bi}
    B_{s-i} = \sigma^{s-i}   .
\end{align}
The first term is of the following form:  
\begin{align}
\label{eq:ai}
    A_i = \sum_{j=1}^i c_j^{(i)} \sigma^j.
\end{align}
The prefactors $c_j^{(i)}$ are given by
 \begin{align}
 \label{eq:Prefactor}
c_j^{(i)} = 
     \left\{
 \begin{aligned}
&\sum_{l=1}^{i-1} c_1^{(l)} \left( \wp_{i-l} - 1 \right),\quad &j=1\\
&c_{j-1}^{(i-1)},\quad &j>1\\
\end{aligned}
\right.
\end{align}
where the first factor is fixed to $c_1^{(1)} = 1$ and with $\wp_s = \left( 1- \left( 1 - p \right)^s \right)^n$.

Before going on, we explain Eq.~(\ref{eq:Tn}) in detail and prove its validity via induction. 
To understand where the decomposition of $T'^{s}$ comes from, it is best to consider a concrete example (e.g., $T'^2$) first. We note that with
\begin{align}
    \sigma^s = \begin{pmatrix}
        (1-p)^s & 0 \\
        1-(1-p)^s & 1
    \end{pmatrix}
\end{align}
we find the following:
\begin{align}
     \begin{pmatrix}
      0 & 1 
  \end{pmatrix}^{\otimes n}
  \sigma
  \begin{pmatrix}
      0  \\
      1 
  \end{pmatrix}^{\otimes n}  
  = \begin{pmatrix}
      0 & 1 
  \end{pmatrix}^{\otimes n}
  \sigma^s
  \begin{pmatrix}
      0  \\
      1 
  \end{pmatrix}^{\otimes n} &= 1  
\end{align}
and
  \begin{align}
  \begin{pmatrix}
      0 & 1 
  \end{pmatrix}^{\otimes n}
  \sigma^s 
  \begin{pmatrix}
      1  \\
      0 
  \end{pmatrix}^{\otimes n} &=  \left( 1 - \left( 1-p\right) ^s \right)^n
\end{align}
Therefore, we can rewrite the term $X_n  \sigma^s  X_n$ by the term $\left(\wp_s -1 \right) X_n$ with $\wp_s = (1-(1-p)^s)^n$, which only depends on the number of parties $n$ and the number of rounds $s$. $T'^2$ can thus be reformulated in the following way: 
\begin{align}
    T'^2 &= \left( \sigma + \sigma X_n  \right) \left( \sigma + \sigma  X_n  \right)  \nonumber \\
    &= \sigma ^2 + \sigma  X_n  \sigma + \left( \left(\wp_1 -1\right) \sigma + \sigma^2 \right) X_n  \sigma^0   \label{eq:T2}
\end{align}
Here, we already find the structure from Eq. (\ref{eq:Tn}). To see that this is not by chance, we additionally calculate $T'^3$ in a similar way:
\begin{align}
    T'^3 &= \left( \sigma^2 + \sigma X_n  \sigma + \left(\wp_1 -1\right) \sigma X_n + \sigma^2  X_n  \right) \left( \sigma + \sigma  X_n  \right) \nonumber \\
    &= \sigma^3   + \sigma  X_n  \sigma^2  + \left( \left(\wp_1 -1\right) \sigma + \left( \sigma \right)^2 \right)  X_n  \sigma  \nonumber \\
    + &\Big\{\Big(( \wp_2 -1) + \left(\wp_1 -1\right)^2 \Big)  \sigma
    + \left( \wp_1 -1 \right)  \sigma^2 
    + \sigma^3 \Big\}
    \!X_N 
    \sigma^0.
\end{align}
This leads again to an expression of the form:
\begin{align}
    T'^s = \sigma^s \sum A_i  X_n  B_{s-i}
\end{align}
with $B_{s-i} = \sigma^{s-i}$.
Now, we can guess the general behavior for a term $T'^s = T'^{s-1} T'$:
\begin{itemize}
    \item Multiplying $T'^{s-1}$ with $\sigma$ leaves the factor $A_i$ untouched and only increases the power of the last term by one.
    \item Multiplying $T'^{s-1}$ with $\sigma X_n$ creates a new term in the sum since $i$ increases by one, i.e., there is a new term with factors $A_s$ and $B_0$. 
    \item The new factor $A_s$ for the last term is given by
    \begin{align}
        A_s = \sum_{j=1}^{s-1} A_j \left( \wp_{s-j} -1 \right) + \sigma^n 
    \end{align}
\end{itemize}
By rewriting the term for the $A_i$ to the form
\begin{align}
    A_i = \sum_{j=1}^i = c_j^{(i)} \sigma^j ,
\end{align}
one finds the prefactors $c_j^{(i)}$ as given in Eq. (\ref{eq:Prefactor}):
\begin{align}
c_j^{(i)} = 
     \left\{
 \begin{aligned}
\sum_{l=1}^{i-1} c_1^{(l)} \left( \wp_{i-l} - 1 \right),\quad j=1\\
c_{j-1}^{(i-1)},\quad j>1\\
\end{aligned}
\right.
\end{align}
with the first factor $c_1^{(1)} = 1$.
\\
\par
Let us prove the equations
\begin{align}
    T'^s &= \sigma^s + \sum_{i = 1}^{s} A_i  X_n  B_{s-i}, \label{eq:IV1} \\
    A_i &= \sum_{j=1}^{i-1} A_j \left(\wp_{i-j} - 1 \right) + \sigma^i  \label{eq:IV2}, \\
    B_i &= \sigma^i \label{eq:IV3}
\end{align}
by induction. 
\begin{proof}
Starting with the base case, we show equality for $s=2$. From Eq.~(\ref{eq:T2}), it follows
\begin{align*}
    T'^2 &=  \sigma^2 + \sigma  X_n  \sigma + \left(\wp_1 -1\right) \sigma X_n 
    + \sigma^2  X_n \nonumber \\
    &= \sigma ^2 + \sigma  X_n  \sigma + \left\{ \left(\wp_1 -1\right) \sigma + \sigma^2 \right\}  X_n \\
    &= \sigma ^2 + A_1  X_n  B_1 + A_2  X_n  B_0  \\
    &= \sigma ^s + \sum_{i=1}^s A_i  X_n  B_{s-i}
\end{align*}
with $A_i$ given in Eq.~(\ref{eq:IV2}) and $B_i$ given in Eq.~(\ref{eq:IV3}):
\begin{align*}
    A_1 &= \sigma,\quad A_2 = \left( \wp_1 -1\right)  \sigma +  \sigma^2, \\
    B_1 &= \sigma, \quad B_0 = 1.
\end{align*}
In the general case, we assume that the equations (\ref{eq:IV1})-(\ref{eq:IV3}) hold for any $s \in \mathbb{N}$. In the induction step, we show that if the statement holds for any $s \in \mathbb{N}$, then $s+1$ follows: 
\begin{align*}
    &T'^{s+1} = (T'^s) T'  \\
    &= \left(  \sigma^s + \sum_{i = 1}^{s} A_i  X_n  B_{s-i} \right) \left( \sigma + \sigma X_n \right)   \\
    &= \sigma^{s+1}  + \sigma^{s+1}  X_n + \sum_{i = 1}^{s} A_i X_n  B_{s-i+1} \\
    &\hspace{2.8cm}+ \sum_{i=1}^s A_i \left( \wp_{s-i+1} - 1\right)  X_n   \\
    &= \sigma^{s+1} 
    + \sum_{i = 1}^{s} A_i  X_n  B_{s-i+1} 
    \\&\hspace{1.2cm}+ \left( \sum_{i=1}^s A_i \left( \wp_{s-i+1} - 1\right) + \sigma^{s+1} \right)  X_n \\
    &= \sigma^{s+1} 
    + \sum_{i = 1}^{s} A_i  X_n  B_{s-i+1} 
    + A_{s+1}  X_n  B_0 \\
    &= \sigma^{s+1} + \sum_{i = 1}^{s+1} A_i  X_n  B_{s-i+1}  \\
    &= T'^{s+1}.
\end{align*}
\end{proof}

\par
To get the probability distribution in the asymptotic limit, we choose an initial configuration and further calculate $T'^{s \rightarrow \infty} \pi^{(0)}$. It turns out that many terms cancel when choosing $\pi^{(0)} = \begin{pmatrix}
      0 & 1 
  \end{pmatrix}^T$, which means that in the initial configuration, all memories are filled. As the goal is to calculate the average number of GHZ measurements per round, we are only interested in the last entry of the stationary distribution $\pi^*$, which gives the probability that all memories are filled (see Eq. (\ref{eq:Prob_L1})). Therefore, we need to calculate 
\begin{align}
     &\begin{pmatrix}
      0 & 1 
  \end{pmatrix}^{\otimes n}
  T'^s 
  \begin{pmatrix}
      0  \\
      1 
  \end{pmatrix}^{\otimes n}
  \nonumber \\&= \begin{pmatrix}
      0 & 1 
  \end{pmatrix}^{\otimes n}
  \left[ \sigma^s + \sum_{i=1}^s A_i X_n B_{s-i} \right] 
  \begin{pmatrix}
      0  \\
      1 
  \end{pmatrix}^{\otimes n}
  \nonumber \\
    &= 1 + \sum_{i=2}^s \sum_{j = 1}^{i-1} c_1^{(j)} \left( \wp_{i-j} -1 \right) \nonumber \\
    &= \sum_{j=1}^s c_1^{(j)} 
\end{align}
By rearranging the sums and considering $\lim_{s \rightarrow \infty}$, the asymptotic result follows:
\begin{align}
\begin{pmatrix}
      0 & 1 
  \end{pmatrix}^{\otimes n}
  T'^s
  \begin{pmatrix}
      0  \\
      1 
  \end{pmatrix}^{\otimes n}  &= 
\sum_{j=1}^\infty c_1^{(j)}  \nonumber 
\\
&= 1 + \sum_{j=1}^\infty c_1^{(j)} \sum_{i = 1}^\infty \left( \wp_i -1 \right)  \nonumber \\ 
\end{align}
\begin{align}
    \Rightarrow    
    \begin{pmatrix}
      0 & 1 
  \end{pmatrix}^{\otimes n}
  T'^s
  \begin{pmatrix}
      0  \\
      1 
  \end{pmatrix}^{\otimes n}  
  &= \frac{1}{1+ \sum_{j=1}^\infty \left( 1-\wp_j  \right)} \nonumber \\
  &= \frac{1}{1+ \sum_{j=1}^\infty \left( 1- \left( 1 - \left(1-p\right)^j \right)^n  \right)} \nonumber \\ &= \langle L_1 \rangle  
  \label{eq:l1}
\end{align}
with the sum over the rounds, here denoted as $j$.

\section{Stationary expectation and dispersion for the simplified model}
\label{sec:ProofSimpModel}

The aim of this section is to derive Eq.~(\ref{eq:ellgen}). Recall that $X_{i,k}$ are identically distributed. Denote  $\langle X_{i,k}\rangle=\mu_k$ and ${\rm Var}[X_{i,k}]=\sigma^2_k$. 

\begin{lemma}
\label{lem:musigma}
The following recurrence relations hold:
\begin{equation}
\label{eq:musigmaprime}
\begin{split}
    \mu_{k+1}&=
    p m+(1-p)\mu_k-l,\\
    \sigma^2_{k+1}&=
    (1-p)^2\sigma^2_k
    +p(1-p)(m-\mu_k).
\end{split}
\end{equation}
\end{lemma}
Note that, unlike other steps of the present analysis, this lemma is a mathematically rigorous statement about a well-defined probabilistic model. The proof will be given later.

Take the limit $k\to\infty$ in Eqs.~(\ref{eq:musigmaprime}) and denote $\mu=\lim_{k\to\infty}\mu_k$ and $\sigma^2=\lim_{k\to\infty}\sigma^2_k$. We have
\begin{equation}
\label{eq:musigma}
\begin{split}
    \mu&=m-\frac{l}{p},\\
    \sigma^2&=
    \frac{1-p}{2-p}
    \frac{l}{p}.
\end{split}
\end{equation}

In order to calculate $\langle\min_i \tilde X_{i,k}\rangle$, see Ineq.~(\ref{eq:EminPositive}), we need to know the distribution of $\tilde X_{i,k}$, not just the expectation value and variance. Numerical simulations show that, for large $k$, even for small $n$ and $m$ it is well approximated by the normal distribution with the expectation $\mu_k$ and the dispersion $\sigma_k^2$ given above. Under this assumption, we can use the known approximation (asymptotics) of the minimum of $n$ identically normally distributed random variables as $n\to\infty$~\cite{OrdStat}:
\begin{equation}
\label{eq:meanmin}
 \langle
 \min
 (\tilde X_{1,k},\ldots,\tilde X_{n,k})
 \rangle
 =\mu_k-\alpha\sigma_k\sqrt{\ln n},  
\end{equation}
and, thus,
\begin{equation}
\label{eq:meanmin}
 \lim_{k\to\infty}
 \langle
 \min(\tilde X_{1,k},\ldots,\tilde X_{n,k})
 \rangle
 =\mu-\alpha\sigma\sqrt{\ln n},  
\end{equation}
where
\begin{equation}
    \alpha=
    \sqrt{2}
    \left(
    1-
    \frac{\ln(4\pi\ln n)-2\gamma}
    {4\ln n}
    \right)
\end{equation}
and $\gamma$ is the Euler-Mascheroni constant.  However, the simulations show that $\alpha=1$ gives a good approximation for a broad range of parameters.

In order to find the maximal  value $l_{\max}$ of $l$ such that Ineq.~(\ref{eq:EminPositive}) holds, we
substitute Eqs.~(\ref{eq:musigma}) into Eq.~(\ref{eq:meanmin}). This gives a quadratic function of $\sqrt{l}$. Its substitution into Ineq.~(\ref{eq:EminPositive}) gives
\begin{equation}
    l_{\max}=
    p m
    \left(
    \sqrt{
    \frac{\alpha^2\beta}{4m}
    \ln n
    +1
    }
    -
    \sqrt{
    \frac{\alpha^2\beta}{4m}
    \ln n
    }
    \right)^2,
\end{equation}
where $\beta=(1-p)/(2-p)$. We take this as an approximation for $\langle L\rangle$ and obtain (\ref{eq:ellgen}).

Let us again summarize the approximations used in this simplified model for the derivation of approximation (\ref{eq:ellgen}) for $\langle L\rangle$:
\begin{itemize}
    \item Substitution of random $\ell$ in the original model by a fixed number $\tilde\ell$ associated with the average value $\langle\ell\rangle$ in the stationary regime.
    
    \item We allow the occupation numbers $\tilde X_i$ to be negative and demand only that their stationary average values are non-negative.

    \item The use of the normal distribution for stationary occupation numbers $\tilde X_i$.

    \item The use of the asymptotic expressions $n\to\infty$ for the case of finite or even small $n$.
\end{itemize}
The first two assumptions are based on the law of large numbers (or neglection of fluctuations) argument, while the third one -- on the central limit theorem argument. Also, these assumptions are justified by simulations. However, it would be interesting to obtain mathematically rigorous justifications.

\begin{proof}[Proof of Lemma~\ref{lem:musigma}]

We have [see Eq.~(\ref{eq:PrYX})] 
\begin{equation}
    \Pr[\tilde Y_{i,k}=y|\tilde X_{i,k-1}=x]=\binom{m-x}{y}p^y(1-p)^{m-x-y},
\end{equation}
$y=0,1,\ldots,m-x$. Denote 
\begin{equation}
\begin{split}
    \mathbb E[\tilde Y_{i,k}
    |\tilde X_{i,k-1}=x]
    &=\sum_{y=0}^{m-x}
    y\Pr[\tilde Y_{i,k}=y|\tilde X_{i,k-1}=x]\\&=p(m-x)
\end{split}
\end{equation}
the conditional expectation of $\tilde Y_{i,k}$ on the condition that $\tilde X_{i,k}$ takes the value $x$ (in the second line we used the expression for the expectation value for the binomial distribution) and
\begin{equation}
\label{eq:EYX}
    \mathbb E[\tilde Y_{i,k}
    |\tilde X_{i,k-1}]
    =p(m-\tilde X_{i,k-1})
\end{equation}
the conditional expectation of $\tilde Y_{i,k}$ conditioned on the random variable $\tilde X_{i,k}$ (informally speaking, regarding $\tilde X$ is a fixed number). This is in accordance with the standard definition of the conditional expectation and $\mathbb E[\tilde Y_{i,k}
    |\tilde X_{i,k-1}]$ is still a random variable. Then
\begin{equation}
\label{eq:EY}
\begin{split}
    \langle Y_{i,k}\rangle
    &=
    \sum_{x=-\infty}^m
    \Pr[\tilde X_{i,k-1}=x]
    \,
    \mathbb E[\tilde Y_{i,k}
    |\tilde X_{i,k-1}=x]
    \\
    &=\big\langle
     \mathbb E[\tilde Y_{i,k}
    |\tilde X_{i,k-1}]
    \big\rangle
        =p(m-\mu_{k-1}).
\end{split}
\end{equation}
Generally,
\begin{equation}
\label{eq:EEcond}
\langle A\rangle=
\big\langle
\mathbb E[A|\tilde X_{i,k-1}]
\big\rangle
\end{equation}
for an arbitrary random variable $A$, which is a general property of the conditional expectation and which we will use.

Since $\tilde Z_{i,k}=\tilde X_{i,k-1}+\tilde Y_{i,k}$ and $\tilde X_{i,k}=\tilde Z_{i,k}-l$,
\begin{equation}
    \langle
    \tilde Z_{i,k}
    \rangle=\mu_{k-1}+(m-\mu_{k-1})p=mp+(1-p)\mu_{k-1}
\end{equation}
and
\begin{equation}
    \mu_k=\mathbb E[\tilde X_{i,k}]=mp+(1-p)\mu_{k-1}-l.
\end{equation}

We have obtained the first formula in Eqs.~(\ref{eq:musigmaprime}). The calculation for the dispersion is as follows:

\begin{equation}
\label{eq:sigmapr}
\begin{split}
    \sigma_k^2
    &\equiv
    {\rm Var}[\tilde X_{i,k}]
    =
    {\rm Var}[\tilde Z_{i,k}]
    \\
    &=
    {\rm Var}[\tilde X_{i,k-1}]
    +
    {\rm Var}[\tilde Y_{i,k}]
    +
    2\,{\rm cov}[\tilde X_{i,k-1},\tilde Y_{i,k}]
    \\
    &=
    \sigma^2
    +
    {\rm Var}[\tilde Y_{i,k}]
    +
    2\,{\rm cov}[\tilde X_{i,k-1},\tilde Y_{i,k}],
\end{split}
\end{equation}
where ${\rm cov}[\tilde X_{i,k-1},\tilde Y_{i,k}]$ denotes the covariance of the two random variables. We have
\begin{equation}
\label{eq:covxy}
    \begin{split}
    &{\rm cov}[\tilde X_{i,k-1},\tilde Y_{i,k}]
    \\&=
    \big\langle
    \big(\tilde X_{i,k-1}-\mathbb \langle\tilde X_{i,k-1}\rangle\big)
    \big(\tilde Y_{i,k}-\mathbb \langle\tilde Y_{i,k}\rangle
    \big)
    \big\rangle
    \\
    &=
    \big\langle
    \mathbb E\big[
    \big(\tilde X_{i,k-1}-
    \langle
    \tilde X_{i,k-1}
    \rangle
    \big)
    \big(\tilde Y_{i,k}-
    \langle
    \tilde Y_{i,k}
    \rangle
    \big)
    \big|\tilde X_{i,k-1}\big]
    \big\rangle
    \\
    &=
    \big\langle
    \big(
    \tilde X_{i,k-1}-
    \langle
    \tilde X_{i,k-1}
    \rangle
    \big)
    \,
    \mathbb E\big[
    (
    \tilde Y_{i,k}-
    \langle \tilde Y_{i,k}
    \rangle
    )
    \big|\tilde X_{i,k-1}\big]
    \big\rangle
    \\
    &=
    \big\langle
    \big(
    \tilde X_{i,k-1}-
    \langle
    \tilde X_{i,k-1}
    \rangle
    \big)\,
    p(\mu_{k-1}-\tilde X_{i,k-1})
    \big\rangle
    \\
    &=
    -p
    \langle
    (\tilde X_{i,k-1}-\mu_{k-1})^2
    \rangle=-p\sigma^2_k.
    \end{split}
\end{equation}
Here we have used Eqs.~ (\ref{eq:EYX})--(\ref{eq:EEcond}). 

Now let us calculate ${\rm Var}[\tilde Y]$:
\begin{equation}
    \begin{split}
    {\rm Var}[\tilde Y_{i,k}]
    &=
    \big\langle
    (\tilde Y_{i,k}-\langle
    \tilde Y_{i,k}
    \rangle)^2
    \big\rangle
    \\
    &=
    \big\langle
    \mathbb E\big[
    (\tilde Y_{i,k}-
    \langle
    \tilde Y_{i,k}
    \rangle)^2
    \big|\tilde X
    \big]
    \big\rangle,
    \end{split}
\end{equation}
where we have again used Eq.~(\ref{eq:EEcond}).
Express now 
\begin{equation}
\begin{split}
    &(\tilde Y_{i,k}-
    \langle 
    \tilde Y_{i,k}
    \rangle)^2
    \\&=
    \left\{
    \big(\tilde Y_{i,k}-\mathbb E[\tilde Y_{i,k}|\tilde X_{i,k-1}]\big)
    +
    \big(\mathbb E[\tilde Y_{i,k}|\tilde X_{i,k-1}]-\langle
    \tilde Y_{i,k}
    \rangle\big)
    \right\}^2
    \\
    &=
    \big(\tilde Y_{i,k}-\mathbb E[\tilde Y_{i,k}|\tilde X_{i,k-1}]\big)^2
    +
    \big(\mathbb E[\tilde Y_{i,k}|\tilde X_{i,k-1}]-\langle
    \tilde Y_{i,k}\rangle\big)^2
    \\&
    +2\big(\tilde Y_{i,k}-\mathbb E[\tilde Y_{i,k}|\tilde X_{i,k-1}]\big)
    \big
    (\mathbb E[\tilde Y_{i,k}|\tilde X_{i,k-1}]-\langle\tilde Y_{i,k}
    \rangle\big)
\end{split}
\end{equation}
and notice
\begin{equation}
    \mathbb E\{(\tilde Y_{i,k}-\mathbb E[\tilde Y_{i,k}|\tilde X_{i,k-1}])^2|\tilde X_{i,k-1}\}=p(1-p)(m-\tilde X_{i,k-1})
\end{equation}
(the variance of the binomial distribution), 
\begin{equation}
\mathbb E[\tilde Y_{i,k}|\tilde X_{i,k-1}]-
\langle\tilde Y_{i,k}
\rangle=p(\mu-\tilde X_{i,k-1}),
\end{equation}
and
\begin{eqnarray}
    &&\mathbb E
    \big\{
    \big(\tilde Y_{i,k}-\mathbb E[\tilde Y_{i,k}|\tilde X_{i,k-1}]\big)
    \big(\mathbb E[\tilde Y_{i,k}|\tilde X_{i,k-1}]-\langle
    \tilde Y_{i,k}
    \rangle
    \big)\big|\tilde X_{i,k-1}
    \big\}
    \nonumber\\
    &&=\!
    \big(\mathbb E[\tilde Y_{i,k}|\tilde X_{i,k-1}]-\langle
    \tilde Y_{i,k}
    \rangle\big)
    \,
    \mathbb E\big\{
    \tilde Y_{i,k}-\mathbb E[\tilde Y_{i,k}|\tilde X_{i,k-1}]
    \big|\tilde X_{i,k-1}\big\}
    \nonumber\\&&=0
\end{eqnarray}
(the last factor is zero).
Hence,
\begin{equation}
\label{eq:dispy}
    \begin{split}
    {\rm Var}[\tilde Y_{i,k}]&=
    \big\langle
    p(1-p)\tilde X_{i,k-1}
    +p^2(\mu-\tilde X_{i,k-1})^2
    \big\rangle
    \\&=
    p(1-p)(m-\mu_{k-1})+
    p^2\sigma^2_{k-1}.
    \end{split}
\end{equation}
Substitution of Eqs.~(\ref{eq:covxy}) and (\ref{eq:dispy}) into Eq.~(\ref{eq:sigmapr}) gives the second formula in Eq.~(\ref{eq:musigmaprime}).

\end{proof}

\end{document}